\documentstyle[aps,epsf]{revtex}
\begin{document}
\addtolength{\baselineskip}{0.1\baselineskip}
\preprint{Submitted to PRA}
\title{The Haroche-Ramsey experiment as a generalized measurement }
\author{W.M. de Muynck \ and A.J.A. Hendrikx}
\address{Theoretical Physics, Eindhoven University
of Technology,\\ P.O.B. 513, 5600 MB Eindhoven, The Netherlands}
\date{\today}
\maketitle
\begin{abstract}
A number of atomic beam experiments, related to the Ramsey experiment
and a recent experiment by Brune et al. \cite{BrHa96},
are studied with respect to the question of complementarity. 
Three different procedures for
obtaining information on the state of the incoming atom are compared.
Positive operator-valued measures are explicitly calculated.
It is demonstrated that, in principle, it is possible to choose the
experimental arrangement so as to admit an
interpretation as a joint non-ideal measurement yielding interference and
``which-way'' information. 
Comparison of the different measurements gives
insight into the question of which information is provided by a
(generalized) quantum mechanical measurement. For this purpose the
subspaces of Hilbert-Schmidt space, spanned by the operators of the
POVM, are determined for different measurement arrangements and
different values of the parameters. \\


\noindent
Pacs number(s): 03.65.Bz, 03.65.Ca
\end{abstract}



\section{Introduction}

In the standard formalism of quantum mechanics a measurement corresponds to
a self-adjoint operator yielding its eigenvalues as measurement results,
with probabilities determined by the expectation values of the projections
on its eigenvectors. It is by now well-known \cite{povm} that this formalism
is too restricted to encompass all possible experiments within the domain of
application of quantum mechanics. Many experiments performed in actual
practice are of the type of {\it generalized measurements }yielding
probabilities determined by the expectation values of so-called positive
operator-valued measures (POVMs) rather than projection-valued ones. The
theory describing this generalization{\normalsize \ is referred to as the 
{\em operational approach.}}

Such generalized measurements are important for two different reasons.
First, from a fundamental point of view they are interesting because they
enable to transcend the bounds imposed by the standard formalism on the
notion of a quantum mechanical measurement. In particular, the projection
postulate does not make sense in a generalized measurement because in
general no projection operators are available. The inapplicability of this
postulate in most realistic measurement procedures indeed demonstrates the
restricted scope of the standard formalism. What is a quantum mechanical
measurement, and which observable is measured by a particular measurement
procedure must be determined by considering in detail the interaction of the
microscopic object and the measuring instrument as a quantum mechanical
process. In the second place generalized measurements are interesting from a
practical point of view because they can yield more information on the state
of the object than is provided by the measurement of a standard observable.

We should distinguish between the determinative
and the preparative aspects of measurement \cite{dMquantph99}. The
former is related to the final state of the {\it measuring instrument, }and
the {\it information on the initial object state that is provided by the
measurement.} The latter refers to the {\it preparation} of the
post-measurement state of the object. Unfortunately, these different aspects
of measurement have been confounded ever since in the Copenhagen
interpretation a measurement has been defined as a preparation of the object
in a final state described by an eigenvector of the measured observable.
It is important to note that most measurement procedures do not satisfy
this criterion.
Due to the ensuing confusion even recently the role of the Heisenberg
uncertainty relations in quantum measurement has been a source of
controversy \cite{StTaCoWa95,ScEnWa95,DuNoRe98}. {\normalsize Whereas Storey
et al.} \cite{StTaCoWa95}{\normalsize \ conclude that ``the principle of
complementarity is a consequence of the Heisenberg uncertainty relation,''
Scully et al.} \cite{ScEnWa95}{\normalsize \ observe that ``The principle
of complementarity is manifest although the position-momentum uncertainty
relation plays no role.'' D\"{u}rr et al. \cite{DuNoRe98} stress that quantum
correlations due to the interaction of object and detector, rather than
``classical'' momentum transfer, enforces the loss of interference in a
`which-way' measurement. In their experiment momentum disturbance is not
large enough to account for the loss of interference if the measurement
arrangement is changed so as to yield `which-way' information. These
diverging statements can easily be reconciled if it is realized that the
Heisenberg inequalities refer to the {\em initial} state of the object, and,
as already stressed a long time ago by Ballentine \cite{Bal70}, do not refer
to the measurement procedure (although, of course, the post-measurement
state of the object will once again satisfy the Heisenberg inequalities if
measurements are performed in this latter state). It should be
realized, however, that in general there need not exist a direct
relation between the determinative properties of a measurement,
yielding information on the initial state of the object, and its
preparative ones, determining what is the final object state.}

The notion of a generalized measurement is able to clarify the confusion
with respect to the role of the Heisenberg uncertainty relations in quantum
measurement \cite{MadM90}. Whereas the standard formalism only allows the
joint measurement of compatible (standard) observables, is the generalized
formalism able to deal with incompatible ones. Moreover, from this formalism
an inequality has been derived (equation (\ref{3}) below) quantifying the
mutual disturbance of the information on the initial state, that is a
consequence of the incompatibility of the 
observables measured jointly, and that, contrary to the Heisenberg
inequality, is a property of the measurement procedure. In recent years
quite a few measurement procedures have been analyzed satisfying the
characteristics of such joint measurements: for instance, eight-port optical
homodyning \cite{WaCa84,YuSha80}, certain {\normalsize neutron interference
experiments \cite{SuRaTu}, Stern-Gerlach experiments \cite{MadM93}, and a
number of Mach-Zehnder interferometric procedures \cite{dMStMa} are of the
generalized type, allowing an interpretation as a joint measurement of
incompatible observables. Presumably also the experiment by D\"{u}rr et al. 
\cite{DuNoRe98} is of this type, the observables measured jointly being
different from the position-momentum pair.}

In this paper we want to consider a recent atomic beam
experiment \cite{BrHa96} from the same point of view, and demonstrate that
also this experiment has a generalized character. In the experiment, to be
referred to in the following as the Haroche-Ramsey experiment, a Rb atom is
sent through three cavities, $R_1,C$ and $R_2$ (figure~\ref{fig1}),
$R_{1}$ and $R_{2}$ being approximately resonant with a particular transition
between two Rydberg states of the atom. Whereas the experiment without
cavity $C$ is a pure interference experiment, already performed by Ramsey
\cite{Ramsey}, the introduction of cavity $C$ provides the possibility to
obtain 
also `which way' information. The visibility of the interference fringes
decreases as the field amplitude $\gamma $ increases, but it only vanishes
in the limit $\gamma \rightarrow \infty .$ So, for finite $\gamma $
information on both interference and path can be obtained. In
order to actually obtain `which way' information, a measurement must be
carried out on the field in cavity $C$ left behind by the atom. The
Haroche-Ramsey experiment is very analogous to the neutron interference
experiments performed by Summhammer et al. \cite{SuRaTu}, in which an
absorber is inserted into one of the paths, the absorber playing an analogous
role as the cavity $C$ field. Consequently the present analysis is similar
to the analysis of the neutron interference experiments performed in
\cite{MuMa90}. 

Recently there has been much interest in the problem of reconstructing the
initial state of the object on the basis of information provided by quantum
measurements \cite{LePfMo98,Ba98,HeMi96,LePaAr95}. Such a reconstruction is
impossible on the basis of a measurement of a single standard observable. As
demonstrated by Vogel and Risken \cite{VoRi89} in quantum tomography this
goal may be achieved by measuring a number of such observables. Generalized
measurements provide the opportunity to obtain a comparable result using one
single measurement arrangement. In general a generalized measurement need
not allow a complete reconstruction of the initial
state. It then is an interesting question which information \ is provided by
the measurement \cite{dM98}. Also this subject will be discussed using the
Haroche-Ramsey experiment as an example. The analysis is particularly suited
to study the question of decoherence of the C field raised by Davidovich et
al. \cite{DaHa96}. In order to experimentally study the decoherence a second
atom is sent through the cavities, yielding information on the state of the
cavity $C$ field as it was at the time of passage. From the experimental
two-atom correlation found in the experiment it was concluded \cite{BrHa96}
that a decoherence effect exists that cannot be explained by loss of photons
from cavity $C$. From an analysis of the experiment as a generalized
measurement it can be seen, however, that such a conclusion cannot be drawn,
because as a measurement of the cavity $C$ field the measurement on the second
atom can only provide non-ideal information on photon number.

The paper is organized as follows. In section \ref{genmeas} the theory of
generalized measurements, and its application to joint non-ideal measurement
of incompatible observables is briefly reviewed. In section
\ref{Haroche-Ramsey} the Haroche-Ramsey experiment and the
Davidovich-Haroche 
experiments are analyzed as generalized measurements. 
In section \ref{Infasp} we demonstrate that the Davidovich-Haroche
experiment is informationally equivalent to a Haroche-Ramsey experiment
in which a measurement of cavity $C$ photon number is performed in
coincidence with a determination of the final state of the atom. In
this section the
decoherence aspects of the Haroche-Ramsey experiment will be dealt with from
the point of view of generalized measurements.
In section \ref{joint}
an alternative measurement procedure for the Haroche-Ramsey experiment is
discussed, that can be interpreted as a joint measurement of incompatible
observables having the complementary character of the ``classical''
double-slit experiments. 

\section{Generalized measurements}
\label{genmeas}

\subsection{Operational approach}

In the operational approach the experimental probabilities
are calculated by treating the interaction between object and measuring
instrument as a quantum mechanical process.
Quantum mechanical measurement results are associated with pointer positions
of the latter. If $\hat{\rho} $ and $\hat{\rho}_{a}$ are the initial
density operators 
of object and measuring instrument, respectively, then the probability of a
measurement result is obtained as the expectation value of the spectral
representation $\{\hat{E}_{m}^{(a)}\}$ of a pointer observable of the
measuring 
instrument in the final state $\hat{\rho}_{f}=\hat{U}\hat{\rho}
\hat{\rho}_{a}\hat{U}^{\dagger },\;\hat{U}=exp(-i\hat{H}T_f/\hbar )$ of the
measurement. Thus, $p_{m}=Tr_{oa}\hat{\rho }_{f}\hat{E}_{m}^{(a)}$.
This quantity can be interpreted as a property of the {\em %
initial} object state, $p_{m}=Tr_{o}\hat{\rho} \hat{M}_{m}$, with
$\hat{M}_{m}=Tr_{a}\hat{\rho}_{a}\hat{U}^{\dagger
}\hat{E}_{m}^{(a)}\hat{U}$. Whereas in the standard formalism quantum 
mechanical probabilities $p_{m}$ are represented by the expectation values
of mutually commuting projection operators ($p_{m}=Tr\hat{\rho}
\hat{E}_{m},\;\hat{E}_{m}^{2}=\hat{E}_{m},\;[\hat{E}_{m},\hat{E}_{m^{\prime
}}]_{-}=\hat{O}$), \ allows the 
generalized formalism to represent these probabilities by expectation values
of operators $\hat{M}_{m}$ that are not necessarily projection
operators, and need 
not commute ($\hat{M}_{m}^{2}\neq
\hat{M}_{m},\;[\hat{M}_{m},\hat{M}_{m^{\prime }}]_{-}\neq \hat{O}$ in 
general). The operators $\hat{M}_{m},\;\hat{O}\leq \hat{M}_{m}\leq
\hat{I},\;\sum_{m}\hat{M}_{m}=\hat{I}$ 
generate a POVM; the observables of the standard formalism are restricted to
those POVMs of which the elements are mutually commuting projection
operators (so-called projection-valued measures (PVM)). 

\subsection{Non-ideal measurements}

{\normalsize In the generalized formalism it is possible to define a
relation of partial ordering between observables, expressing that the
measurement represented by one POVM can be interpreted as a non-ideal
measurement of another one }\cite{MadM90}{\normalsize. Thus,
a POVM $\{\hat{R}_{m}\}$ is representing a {\em non-ideal} measurement of the
(generalized or standard) observable $\{\hat{M}_{m^{\prime }}\}$ if the
following relation holds between the elements of the POVMs: 
\begin{equation}
\hat{R}_{m}=\sum_{m^{\prime }}\lambda _{mm^{\prime }}\hat{M}_{m^{\prime
}},\;\lambda _{mm^{\prime }}\geq 0,\;\sum_{m}\lambda _{mm^{\prime }}=1.
\label{1} 
\end{equation}
The matrix $(\lambda _{mm^{\prime }})$ is the non-ideality matrix. It is a
so-called {\em stochastic} matrix. Its elements $\lambda _{mm^{\prime }}$
can be interpreted as conditional probabilities of finding measurement
result }$m${\normalsize \ if an ideal measurement had yielded measurement
result }$m^{\prime }${\normalsize. In case of an ideal measurement the
non-ideality matrix $(\lambda _{mm^{\prime }})$ reduces to the unit matrix $%
(\delta_{mm^{\prime }})$. }

{\normalsize Non-ideality relations of type (\ref{1}) are well-known from
the theory of transmission channels in the classical theory of stochastic
processes} \cite{McEl77}{\normalsize , where the non-ideality matrix
describes the crossing of signals between subchannels. It should be noted,
however, that, notwithstanding the classical origin of the latter subject,
the non-ideality relation (\ref{1}) may be of a {\em quantum mechanical}
nature. Shannon's channel capacity can be used also in this latter case to
quantify the deviation of a non-ideal measurement from ideality
\cite{MadM90}. Another 
useful measure of the departure of a non-ideality matrix from the unit
matrix is the average row entropy of the non-ideality matrix $(\lambda
_{mm^{\prime }})$, 
\begin{equation}
J_{(\lambda )}=-\frac{1}{N}\sum_{mm^{\prime }}\lambda _{mm^{\prime }}\ln 
\frac{\lambda _{mm^{\prime }}}{\sum_{m^{\prime \prime }}\lambda _{mm^{\prime
\prime }}},  \label{rowent}
\end{equation}
which (restricting to square $N\times N$ matrices) satisfies the following
properties: 
\[
\begin{array}{l}
0\leq J_{(\lambda )}\leq \ln N, \\ 
J_{(\lambda )}=0\mbox{ \rm if }\lambda _{mm^{\prime }}=\delta _{mm^{\prime
}}, \\ 
J_{(\lambda )}=\ln N\mbox{ \rm if }\lambda _{mm^{\prime }}=\frac{1}{N}.
\end{array}
\]
Hence, the quantity $J_{(\lambda )}$ vanishes in case of an ideal
measurement of observable $\{\hat{M}_{m^{\prime }}\}$, and obtains its maximal
value if the measurement is uninformative (i.e. does not yield any
information on the observable measured non-ideally; this is the case if in
(\ref{1}) }$\lambda _{mm^{\prime }\text{ }}$ is independent of $m^{\prime }$,
and, hence, $\hat{R}_{m}$=$\alpha _{m}\hat{I},${\normalsize \ }$\alpha _{m}$
constants 
{\normalsize for all }$m${\normalsize ) due to maximal disturbance of the
measurement results. In the following we shall use the non-ideality measure (%
\ref{rowent}).}

\subsection{Joint non-deal measurement of incompatible observables}
\label{jointnonid}
Within the {\normalsize generalized formalism of POVMs it is possible to
extend the notion of quantum mechanical measurement to the joint measurement
of two (generalized) observables. Such a
measurement is required to yield a {\em bivariate} joint probability
distribution $ p_{mn}$, satisfying $p_{mn}\geq 0,\sum_{mn}p_{mn}=1$.
Here $m$ and $n$ label 
the possible values of the two observables measured jointly, corresponding
to pointer positions of two different pointers (one for each observable)
being jointly read for each individual preparation of an object. It is
assumed that, analogous to the case of single measurement, the probabilities 
$p_{mn}$ of finding the pair $(m,n)$ are represented in the formalism by the
expectation values $Tr\hat{\rho} \hat{R}_{mn}$ of a bivariate POVM
$\{\hat{R}_{mn}\},\;\hat{R}_{mn}%
\geq \hat{O},\;\sum_{mn}\hat{R}_{mn}=\hat{I}$ in the initial state\
}$\hat{\rho} $ of 
the object. Then the marginal probabilities $\{\sum_{n}p_{mn}\}$ and $%
\{\sum_{m}p_{mn}\}$ are expectation values of POVMs
$\{\hat{M}_{m}=\sum_{n}\hat{R}_{mn}%
\} $ and $\{\hat{N}_{n}=\sum_{m}\hat{R}_{mn}\}$, respectively, which
correspond to the 
(generalized) observables measured jointly. A measurement,
represented by a bivariate POVM $\{\hat{R}_{mn}\}$, can be interpreted
as a {\em %
joint non-ideal} measurement of the observables $\{\hat{E}_{m}\}$ and
$\{\hat{F}_{n}\}$ 
if the marginals $\{\sum_{n}\hat{R}_{mn}\}$ and
$\{\sum_{m}\hat{R}_{mn}\}$ of the 
bivariate POVM $\{\hat{R}_{mn}\}$ represent {\em non-ideal} measurements of
observables $\{\hat{E}_{m}\}$ and $\{\hat{F}_{n}\}$, respectively. Then, in
accordance with (\ref{1}) two non-ideality matrices
$(\lambda_{mm^{\prime }})$ and $(\mu_{nn^{\prime }})$ exist,
such that 
\begin{equation}
\begin{array}{l}
\sum_{n}\hat{R}_{mn}=\sum_{m^{\prime }}\lambda _{mm^{\prime
}}\hat{E}_{m^{\prime 
}},\;\lambda _{mm^{\prime }}\geq 0,\;\sum_{m}\lambda _{mm^{\prime }}=1, \\ 
\sum_{m}\hat{R}_{mn}=\sum_{n^{\prime }}\mu _{nn^{\prime }}\hat{F}_{n^{\prime
}},\;\mu _{nn^{\prime }}\geq 0,\;\sum_{n}\mu _{nn^{\prime }}=1.
\end{array}
\label{2}
\end{equation}
It is possible that $\{\hat{E}_{m}\}$ and $\{\hat{F}_{n}\}$ are
standard observables. It 
will be demonstrated in the following sections that the Haroche-Ramsey
experiment satisfies the joint measurement scheme given above.

{\normalsize If $\{\hat{E}_{m}\}$ and $\{\hat{F}_{n}\}$ are standard
observables, then 
the non-idealities expressed by the non-ideality matrices $(\lambda
_{mm^{\prime }})$ and $(\mu _{nn^{\prime }})$ can be proven \cite{MadM90} to
satisfy the characteristic traits of the type of complementarity that is due
to mutual disturbance of measurement results (or inaccuracy) in a joint
measurement of incompatible observables. 
Quantifying the non-idealities of the non-ideality matrices $(\lambda
_{mm^{\prime }})$ and $(\mu _{nn^{\prime }})$ by the average row entropy (%
\ref{rowent}), it can be demonstrated \cite{MadM90} that for a joint
non-ideal measurement of two standard observables
$\hat{A}=\sum_{m}a_{m}\hat{E}_{m}$ and 
$\hat{B}=\sum_{n}b_{n}\hat{F}_{n}$, with eigenvectors $|a_{m}\rangle $ and $%
|b_{n}\rangle $, respectively, the non-ideality measures $J_{(\lambda )}$
and $J_{(\mu )}$ obey the following inequality: 
\begin{equation}
J_{(\lambda )}+J_{(\mu )}\geq -2\ln \{max_{mn}|\langle a_{m}|b_{n}\rangle
|\}.  \label{3}
\end{equation}
It is evident that (\ref{3}) is a nontrivial inequality (the right-hand side
unequal to zero) if the two observables $\hat{A}$ and $\hat{B}$ are
incompatible in the 
sense that the operators do not commute. Contrary to the Heisenberg
inequality }$\Delta \hat{A}\Delta \hat{B}\geq \frac{1}{2}\mid Tr\rho \lbrack
\hat{A},\hat{B}]_{-}\mid $,{\normalsize \ inequality (\ref{3}) does not
refer to the 
preparation of the initial state but exclusively to the measurement
process. Inequality (\ref{3}) should also be clearly distinguished from
the entropic uncertainty 
relation} \cite{Deutsch,Partovi,Kraus87,MaUf88} for the standard observables 
$\hat{A}=\sum_{m}a_{m}\hat{E}_{m}$ and $\hat{B}=\sum_{n}b_{n}\hat{F}_{n}$, 
\begin{equation}
H_{\{\hat{E}_{m}\}}(\hat{\rho} )+H_{\{\hat{F}_{n}\}}(\hat{\rho} )\geq
-2\ln \{max_{mn}|\langle a_{m}|b_{n}\rangle |\},  \label{10}
\end{equation}
in which $H_{\{\hat{E}_{m}\}}(\hat{\rho} )=-\sum_{m}p_{m}\ln
p_{m},\;p_{m}=Tr\hat{\rho} \hat{E}_{m}$ 
(and analogously for $\hat{B}$). Inequality (\ref{10}), although quite
similar to inequality {\normalsize (\ref{3})}, should be compared to
the Heisenberg 
inequality, expressing a property of the initial state $\hat{\rho} $,
to be tested 
by means of separate measurements of observables $\{\hat{E}_{m}\}$ and
$\{\hat{F}_{n}\}$. 

\subsection{Informational aspects\label{genmeasinfo}}

The operators $\hat{M}_{m}$ of a POVM span a subspace ${\cal H}_{\{
\hat{M}_{m}\} }$ of the linear space of bounded operators. For
finite-dimensional systems this can be the Hilbert-Schmidt space ${\cal H}%
_{HS}$ having inner product $Tr \hat{A}^{\dagger }\hat{B}.$ If the
operators $\hat{M}_{m}$ are 
linearly independent they constitute a (generally non-orthogonal) basis
of the subspace. 
Within subspace ${\cal H}_{\{ \hat{M}_{m}\} }$ a
bi-orthogonal system is defined by \cite{dM98}

\begin{equation}
\hat{M}_{m^{\prime }}^{\prime }=\sum_{m}\beta _{m^{\prime
}m}\hat{M}_{m},Tr \hat{M}_{m}\hat{M}_{m^{\prime }}^{\prime }=\delta
_{mm^{\prime }},\;\beta _{m^{\prime }m} \hbox{\rm real}. 
\label{biorth}
\end{equation}

In general the set of Hermitian operators $\hat{M}_{m^{\prime
}}^{\prime }$ constitutes 
another non-orthogonal basis of ${\cal H}_{\{ \hat{M}_{m}\} }$. The
information the measurement of POVM $\{\hat{M}_{m}\}$ yields on the
initial 
state $\hat{\rho} $ of the object can be represented by the projection
$\hat{\rho} 
_{\{\hat{M}_{m}\}}={\cal P}_{\{ \hat{M}_{m}\} }\hat{\rho} $
of $\hat{\rho} $ onto ${\cal H }_{\{ \hat{M}_{m}\} }.$ This
projection is given by 

\begin{equation}
{\cal P}_{\{ \hat{M}_{m}\} }\hat{\rho} =\sum_{m}(Tr
\hat{M}_{m}^{\prime }\hat{\rho} 
)\hat{M}_{m}=\sum_{m}(Tr \hat{M}_{m}\hat{\rho} )\hat{M}_{m}^{\prime }.
\label{rho_projected} 
\end{equation}

For complete measurements we have ${\cal P}_{\{ \hat{M}_{m}\}
}\hat{\rho} =\hat{\rho} .$ 
Measurements are less complete as the subspace ${\cal H}_{\{
\hat{M}_{m}\} }$has a smaller dimension. It should be noted that,
contrary 
to an assertion made in \cite{dM98}, for incomplete measurements $\hat{\rho}
_{\{\hat{M}_{m}\}}$ need not be a density operator, even though $Tr\hat{\rho}
_{\{\hat{M}_{m}\}}=1.$ In general $\hat{\rho}_{\{\hat{M}_{m}\}}$ is not
a non-negative 
operator if the object Hilbert space has dimension greater than $2$ (in the
Appendix it is proven that for two-dimensional Hilbert spaces $\hat{\rho}
_{\{\hat{M}_{m}\}}$ is non-negative). Hence,
$\hat{\rho}_{\{\hat{M}_{m}\}}$ should be 
compared to a description of a quantum state by means of the Wigner
distribution, yielding $Tr\hat{\rho}
\hat{A}=Tr\hat{\rho}_{\{\hat{M}_{m}\}}\hat{A}$ for all operators $%
\hat{A}\in {\cal H}_{\{ \hat{M}_{m}\} },$ but not containing any
information on the part of $\hat{\rho}$ that is in the orthogonal complement
of ${\cal H}_{\{\hat{M}_{m}\} }.$ In principle the subspace ${\cal
H}_{\{ \hat{M}_{m}\} }$ completely 
determines the information on the initial density operator $\hat{\rho}$ that
can be retrieved by a measurement of POVM $\{\hat{M}_m\}$.

\section{Atomic beam interference experiments}
\label{Haroche-Ramsey} 

\subsection{The Ramsey experiment}

In the Ramsey experiment \cite{BrHa96} a beam of Rb atoms is sent
through two identical cavities $R_{1}$ and $R_{2}$. The relevant Hilbert
space ${\cal H}$ of a Rb atom is spanned by the orthogonal state vectors $%
\left| e\right\rangle $ and $\left| g\right\rangle .$ These correspond to 
circular Rydberg levels with principal quantum numbers $n=51$ and $n=50,$
respectively (transition frequency
$\omega_{0}=\omega_{e}-\omega_{g}=321\ast 10^{9}$ $rad/s$). The
frequency of the classical microwave 
fields in the cavities is denoted by $\omega $, its amplitude by
$\Omega$ (the Rabi frequency), and
the time needed by an atom to pass one cavity by $T$.\ The unitary
transformation $\hat{U}_{i}$ describing the evolution of the state of a
Rb atom 
while passing cavity $R_{i}$ between $t=t_{i}$ and $t_i+T$ is given in
the $\{ \left| 
e\right\rangle ,\left| g\right\rangle \} $-representation by the
matrix 
\begin{equation}
\hat{U}_{i}=\left( 
\begin{array}{ll}
e^{-i\left( \frac{\nu }{2}+\omega _{e}\right) T}S_{1} & e^{-i\left( \frac{%
\nu }{2}+\omega _{e}\right) T-i\omega t_{i}}S_{2} \\ 
e^{i\left( \frac{\nu }{2}-\omega _{g}\right) T+i\omega t_{i}}S_{2} & 
e^{i\left( \frac{\nu }{2}-\omega _{g}\right) T}S_{1}^{\ast }
\end{array}
\right)  \label{U}
\end{equation}
where $S_{1}=\cos \frac{aT}{2}+\frac{i\nu }{a}\sin \frac{aT}{2}$, $S_{2}=%
\frac{-i\Omega}{a}\sin \frac{aT}{2}$ , $a=\sqrt{\nu ^{2}+\Omega^{2}}$,
 and $\nu
=\omega -\omega _{0}$ the detuning parameter. A derivation of
(\ref{U}) can be found in Ramsey
\cite{Ramsey} and Paul \cite{Paul91}. For all values of the parameters
we have $|S_1|^2+|S_2|^2=1$. The Rb atom is said to undergo a 
$\frac{\pi }{2}$ pulse in cavity $R_{i}$ if
$|S_{1}|^{2}=|S_{2}|^{2}=\frac{1}{2}$. We shall introduce a parameter
$\delta=|S_{1}|^{2}-|S_{2}|^{2}$, quantifying experimental deviation
from the $\frac{\pi }{2}$ pulse condition. Note that satisfaction of
this latter condition does not imply $\nu =0.$ The phase factor $e^{-i\omega
t_{i}}$ in (\ref{U}) takes into account the phase of the microwave
field at the moment the atom enters the cavity.

Let $\left| \psi _{in}\right\rangle $ be the initial state of the atom. If
the standard observable $\{ \left| e\right\rangle \left\langle e\right|
,\left| g\right\rangle \left\langle g\right| \} $ is measured after
the atom has passed cavity $R_{1},$ then the probabilities $p_{e}$ and $p_{g}$ can be
related to the initial state by means of the equalities 
\begin{equation}
\begin{array}{c}
p_{e}=\left\langle \psi _{in}\right| \left( \hat{U}_{1}^{\dagger }\left|
e\right\rangle \left\langle e\right| \hat{U}_{1}\right) \left| \psi
_{in}\right\rangle =\left\langle \psi _{in}\right| \hat{P}_{+}\left| \psi
_{in}\right\rangle , \\ 
p_{g}=\left\langle \psi _{in}\right| \left( \hat{U}_{1}^{\dagger }\left|
g\right\rangle \left\langle g\right| \hat{U}_{1}\right) \left| \psi
_{in}\right\rangle =\left\langle \psi _{in}\right| \hat{P}_{-}\left| \psi
_{in}\right\rangle .
\end{array}
\label{R1}
\end{equation}
Due to the unitarity of $\hat{U}_{1}$ this yields PVM $\{
\hat{P}_{+},\hat{P}_{-}\} $ as the POVM of this experiment, with 
\begin{equation}
\hat{P}_{+}=\left| p_{+}\right\rangle \left\langle p_{+}\right|
=\hat{I}-\hat{P}_{-}\text{ };%
\text{ }|p_{+}\rangle =\left( 
\begin{array}{c}
S_{1}^{\ast } \\ 
S_{2}^{\ast }e^{i\omega t_{1}}
\end{array}
\right) ,  \label{path_obs}
\end{equation}
in which $|p_{+}\rangle $ equals $\hat{U}_{1}^{\dagger }|e\rangle $ up
to a phase 
factor$.$ Because of the analogy with neutron interference experiments
\cite{MuMa90} the observable $\{ \hat{P}_{+},\hat{P}_{-}\} $
will be referred to 
as the {\em path} observable, even though {\normalsize here the paths are
not trajectories in configuration space (as it is in the double-slit
experiment and the neutron interference experiments) but in the Hilbert
space }${\cal H}${\normalsize \ of the internal states of the Rb atom.
Mathematically this does not constitute a difference, however. }

The observable $\{ \hat{P}_{+},\hat{P}_{-}\} $ is dependent on the initial
phase of the microwave field. For this reason it is not allowed to ignore
this phase if the experiment is intended to yield a measurement of the
initial state of the atom. If the atoms are prepared at random phases, then
a measurement of $\{ \left| e\right\rangle \left\langle e\right|
,\left| g\right\rangle \left\langle g\right| \} ,$ performed
immediately after the atom has passed cavity $R_{1},$ will yield
probabilities obtained from (\ref{R1}) by phase averaging. The corresponding
POVM $\{ \overline{\hat{P}_{+}},\overline{\hat{P}_{-}}\} $
is found according to

\begin{eqnarray*}
\overline{\hat{P}_{+}} &=&|S_{1}|^{2}\left| e\right\rangle \left\langle
e\right| +|S_{2}|^{2}\left| g\right\rangle \left\langle g\right| , \\
\overline{\hat{P}_{-}} &=&|S_{2}|^{2}\left| e\right\rangle \left\langle
e\right| +|S_{1}|^{2}\left| g\right\rangle \left\langle g\right| ,
\end{eqnarray*}
and, hence, represents a non-ideal measurement of PVM $\{\left|
e\right\rangle \left\langle e\right| ,\left| g\right\rangle \left\langle
g\right| \}$ in the sense of (\ref{1}). Note that $\{
\overline{\hat{P}_{+}},\overline{\hat{P}_{-}}\} $ 
is uninformative in case the $\frac{\pi }{2}$ pulse condition is satisfied,
since then its expectation values do not yield any information on $\left|
\psi _{in}\right\rangle $.

The Ramsey set-up consists of two cavities $R_{i},i=1,2$, entered by the
atom at times\ $t_{i}$ (with $t_2= t_1+T+\tau,\;\tau>0$). If the
initial state of the Rb atom is 
\begin{equation}
\left| \psi _{in}\right\rangle =\alpha \left| e\right\rangle +\beta \left|
g\right\rangle ,
\label{in} 
\end{equation}
then the final state at time $t_{2}+T$ is: 
\[
\left| \psi _{f}\right\rangle =\left[ \alpha \left( E+F\right) +\beta \left(
G+H\right) \right] \left| e\right\rangle +\left[ \alpha \left( J+K\right)
+\beta \left( L+M\right) \right] \left| g\right\rangle 
\]
where we have used the abbreviations 
\begin{equation}
\begin{tabular}{ll}
$E=S_{1}^{2}e^{-i\left( \nu +2\omega _{e}\right) T-i\omega _{e}\tau }$ & $%
J=S_{1}S_{2}e^{i\nu \tau }e^{i\left( \nu -2\omega _{g}\right) T-i\omega
_{g}\tau }{}^{+i\omega t_{1}}$ \\ 
$F=S_{2}^{2}e^{-i\nu \tau }e^{-i\left( \nu +2\omega _{e}\right) T-i\omega
_{e}\tau }$ & $K=S_{1}^{\ast }S_{2}e^{i\left( \nu -2\omega _{g}\right)
T-i\omega _{g}\tau }{}^{+i\omega t_{1}}$ \\ 
$G=S_{1}S_{2}e^{-i\left( \nu +2\omega _{e}\right) T-i\omega _{e}\tau
-i\omega t_{1}}$ & $L=S_{2}^{2}e^{i\nu \tau }e^{i\left( \nu -2\omega
_{g}\right) T-i\omega _{g}\tau }{}$ \\ 
$H=S_{1}^{\ast }S_{2}e^{-i\nu \tau }e^{-i\left( \nu +2\omega _{e}\right)
T-i\omega _{e}\tau -i\omega t_{1}}$ & $M=S_{1}^{\ast 2}e^{i\left( \nu
-2\omega _{g}\right) T-i\omega _{g}\tau }.$%
\end{tabular}
\label{constants}
\end{equation}

In the Ramsey experiment the standard observable $\{ 
\left| e\right\rangle \left\langle e\right| ,\left| g\right\rangle
\left\langle g\right| \} $ is measured after the atom has passed
cavity $R_{2}$. The corresponding probabilities $p_{e}$ and $p_{g}$ can
be related to the initial state according to
\[
\begin{array}{c}
p_{e}=\left\langle \psi _{f}\right. \left| e\right\rangle \left\langle
e\right| \left. \psi _{f}\right\rangle =\left\langle \psi _{in}\right|
\hat{Q}_{e}\left| \psi _{in}\right\rangle , \\ 
p_{g}=\left\langle \psi _{f}\right. \left| g\right\rangle \left\langle
g\right| \left. \psi _{f}\right\rangle =\left\langle \psi _{in}\right|
\hat{Q}_{g}\left| \psi _{in}\right\rangle,
\end{array}
\]
yielding 
\begin{equation}
\hat{Q}_{e}=\left| q_{e}\right\rangle \left\langle q_{e}\right|
=\hat{I}-\hat{Q}_{g}\text{ ; } \left| q_{e}\right\rangle =\left( 
\begin{array}{c}
\left( E^{\ast }+F^{\ast }\right) \\ 
\left( G^{\ast }+H^{\ast }\right)
\end{array}
\right).  \label{interf_obs}
\end{equation}
The observable $\{ \hat{Q}_{e},\hat{Q}_{g}\} $ can be
interpreted as the 
quantum mechanical observable measured in the Ramsey experiment if the
initial phase of the microwave field is well-defined. It is
easily verified that $\langle q_{e}|q_e\rangle =1$.
So $\hat{Q}_{e}$ and $\hat{Q}_{g}$ are projections and $\{
\hat{Q}_{e},\hat{Q}_{g}\} $ is 
a PVM. This PVM will be referred to as the {\em interference} observable,
because, provided the atom velocity is sufficiently well-defined, its
expectation values in the state $\left| \psi _{in}\right\rangle $ exhibit
interference fringes if $\omega $ is varied. It is easily verified that if
the $\frac{\pi }{2}$ pulse condition $\delta=0$ is
satisfied, and the detuning parameter $\nu $ is taken to be zero, then the
interference observable reduces to PVM $\{ \left| e\right\rangle
\left\langle e\right| ,\left| g\right\rangle \left\langle g\right| \}
. $ This is in agreement with the fact that under these conditions the
Ramsey setup just interchanges the roles of the $e$ and $g$ states. In
general the standard observables $\{
\hat{P}_{+},\hat{P}_{-}\} $ and $\{ 
\hat{Q}_{e},\hat{Q}_{g}\} $ are incompatible. 

Also $\{
\hat{Q}_{e},\hat{Q}_{g}\} $ 
is dependent on the initial phase of the microwave field. Averaging over
this phase yields the POVM $\{
\overline{\hat{Q}_{e}},\overline{\hat{Q}_{g}}%
\}, $%
\begin{eqnarray*}
\overline{\hat{Q}_{e}} &=&|S_{1}^{2}+S_{2}^{2}e^{-i\nu \tau }|^{2}\left|
e\right\rangle \left\langle e\right| +|S_{2}|^{2}|S_{1}+S_{1}^{\ast
}e^{-i\nu \tau }|^{2}\left| g\right\rangle \left\langle g\right| , \\
\overline{\hat{Q}_{g}} &=&\hat{I}-\overline{\hat{Q}_{e}},
\end{eqnarray*}
which, once again, is a non-ideal measurement of $\{ \left|
e\right\rangle \left\langle e\right| ,\left| g\right\rangle \left\langle
g\right| \} $ (for $\nu =\delta=0$ it
even is an ideal one). It is important to note that, nevertheless, the
expectation values of $\{
\overline{\hat{Q}_{e}},\overline{\hat{Q}_{g}}\} $ 
exhibit interference fringes if $\omega $ is varied. When in (\ref{in})
$\beta =0$ (as was satisfied in the experiments that have actually been
carried out), then the 
expectation values of $\hat{Q}_{e}$ and $\overline{\hat{Q}_{e}}$
coincide. For this 
special case it is possible to analyze the Ramsey experiment in terms of the
standard formalism, even if the actually performed experiment is a
generalized one, described by POVM $\{
\overline{\hat{Q}_{e}},\overline{ \hat{Q}_{g}}\} $.

Contrary to the phase-averaged experiment, in case of a
well-defined initial phase of the field the measurement performed after
the atom left $ R_{1}$ is incompatible with the one performed after
$R_{2}$ if $\delta\neq \pm 1$. The two
measurement arrangements are complementary. The experiments are analogous to
double-slit experiments in which it either is directly measured which slit a
particle has passed through (`which way' or path measurement), or the
interference 
pattern is measured after the two partial beams have been allowed to
interfere (interference experiment). The quantity $\nu \tau $ is the
relative phase shift of the partial beams.

In standard quantum mechanics complementarity is interpreted as mutual
exclusiveness of information, caused by the impossibility of having
both experimental arrangements simultaneously. 
Of the two incompatible PVMs $\{{\hat{P}_{+}},{
\hat{P}_{-}}\} $ and $\{ {\hat{Q}_{e}},{ \hat{Q}_{g}}\} $ either one or
the other can be measured. 
In the following we shall discuss a measurement arrangement
that is intermediate between the two arrangements considered above, viz. the
experiment reported by Brune et al. \cite{BrHa96}, to be referred to as the
Haroche-Ramsey experiment. As a result the experiment may yield information
on both the path and the interference observable. 

\subsection{The Haroche-Ramsey experiment}

In the Haroche-Ramsey experiment a third cavity $C$ is placed between
cavities $R_{1}$ and $R_{2}$, storing a coherent field $\left| \gamma
\right\rangle $. The transition frequency $\omega _{0}$ of the Rb atom and
the frequency of the cavity $C$ field are chosen to be off-resonance,
so there is no exchange of energy when the atom passes $C$. Contrary to the
microwave fields in cavities $R_1$ and $R_2$ the cavity $C$ field is
treated quantum mechanically. The field in cavity $C\;$
merely undergoes a phase shift $\Phi $ (single atom index effect) which
depends on the state of the Rb atom in the following way
\cite{BrHa96}: 
\begin{equation}
| e\rangle \otimes | \gamma \rangle \stackrel{C}{%
\rightarrow }| e\rangle \otimes | \gamma e^{i\Phi
}\rangle \text{ };\text{ }| g\rangle \otimes | \gamma
\rangle \stackrel{C}{\rightarrow }| g\rangle \otimes |
\gamma e^{-i\Phi }\rangle.  \label{transformatie C}
\end{equation}

The states $| \gamma e^{i\Phi }\rangle $ and $| \gamma
e^{-i\Phi }\rangle $ are also coherent states. This yields the
following unitary transformation $\hat{U}_{C}$ describing the evolution of the
atom-field system when the atom is going from $R_{1}$ to $R_{2}$ :
\begin{equation}
\hat{U}_{C}=e^{-i\omega _{e}\tau }\left| e\right\rangle \left\langle e\right|
\otimes e^{i\Phi \hat{a}^{\dagger }\hat{a}}+e^{-i\omega _{g}\tau
}\left| g\right\rangle 
\left\langle g\right| \otimes e^{-i\Phi \hat{a}^{\dagger }\hat{a}},
\end{equation}
where $\hat{a}^{\dagger }$ and $\hat{a}$ are the photon creation and
annihilation 
operators of the cavity $C$ field mode. For the initial state $\left| \Psi
_{in}\right\rangle =\left[ \alpha \left| e\right\rangle +\beta \left|
g\right\rangle \right] \otimes \left| \gamma \right\rangle $ of the combined
atom-field system we get as final state:

\begin{equation}
\left| \Psi _{f}\right\rangle =\left\{ 
\begin{array}{ll}
\text{ \ }\left| e\right\rangle \otimes & \left[ \left( \alpha E+\beta
G\right) | \gamma e^{i\Phi }\rangle +\left( \alpha F+\beta
H\right) | \gamma e^{-i\Phi }\rangle \right] \\ 
+\left| g\right\rangle \otimes & \left[ \left( \alpha J+\beta L\right)
| \gamma e^{i\Phi }\rangle +\left( \alpha K+\beta M\right) |
\gamma e^{-i\Phi }\rangle \right] ,
\end{array}
\right.  \label{final state}
\end{equation}
constants $E,G,etc.$ being given by (\ref{constants}).

After the atom has passed cavity $C\;$the field is containing path
information
that can be retrieved by a measurement of a well-chosen observable of the
field. In the standard formalism this information is usually analyzed in
terms of the inner product $\langle \gamma e^{i\Phi }| \gamma e^{-i\Phi
}\rangle $ of the field states $| \gamma e^{i\Phi }\rangle $
and $| \gamma e^{-i\Phi }\rangle $, determining their
distinguishability. The possibility of interference is seen as a consequence
of the indistinguishability of the paths. How distinguishable the paths are,
depends on the values of the parameters $\gamma $ and $\Phi $. If the states
are identical, then the paths are completely indistinguishable. Ignoring the
possibility $\Phi =m\pi $ $(m\in {Z\!\!\!Z}),$ this obtains if $\gamma =0.$ In
this case the experiment cannot yield any path information. Complete
distinguishability, corresponding to maximal path information, obtains if
the field states are orthogonal. This only obtains in the limit $\gamma
\rightarrow \infty .$ In the next sections this analysis will be
corroborated on the basis of the generalized formalism.

Whereas for the limiting values of $\gamma ,$ considered above, the standard
formalism is sufficient, is the generalized formalism necessary for
experiments corresponding to intermediate values $0<\gamma <\infty $. This
already holds true if no measurement of the cavity $C$ field is carried out
at all. Thus, putting $\langle \Psi _{f}$ $\left| e\right\rangle
\left\langle e\right| \Psi _{f}\rangle =\langle \psi _{in}|\hat{R}_{e}\left| \psi
_{in}\right\rangle $ (and analogously for $g)$ we find the POVM $%
\{\hat{R}_{e},\hat{R}_{g}\}$ of the Haroche-Ramsey measurement from (\ref{final state}).
Restricting to $\nu =\delta=0$ we get
%
\begin{equation}
\hat{R}_{e}=\frac{1}{2}\left( 
\begin{array}{cc}
1-C_{1} & e^{-i\omega t_{1}}C_{2} \\ 
e^{i\omega t_{1}}C_{2} & 1+C_{1}
\end{array}
\right) ,\text{ }\hat{R}_{g}=\hat{I}-\hat{R}_{e},  \label{ReRg}
\end{equation}
in which 
\begin{equation}
C_{1}+iC_{2}=\langle \gamma e^{i\Phi }|\gamma e^{-i\Phi }\rangle =e^{-\gamma
^{2}\left( 1-e^{-2i\Phi }\right) }.  \label{C}
\end{equation}
It is easily verified that, unless $C_{1}=\pm 1,C_{2}=0,$
$\{\hat{R}_{e},\hat{R}_{g}\}$ is not a PVM. 
Even in the limit $\gamma \rightarrow \infty $ (corresponding to $%
C_{1}=C_{2}=0)$ (\ref{ReRg})\ is a POVM, be it an uninformative one. This is
consistent with complementarity in the sense that in this limit no
information on the interference observable is obtained, path
information being obtainable by a measurement 
of an observable of the cavity $C$ field in the final state of this field.

Since {\normalsize the operators of POVM
}$\{\overline{\hat{R}_{e}},\overline{\hat{R}_{g}}\},$ obtained from
(\ref{ReRg})\ by phase-averaging, are{\normalsize \ 
diagonal, the Haroche-Ramsey experiment is just a non-ideal version of the 
Ramsey experiment if the initial phase of the microwave field is random. The
non-ideality measure (\ref{rowent}) is then given as} 
\begin{equation}
J^{\overline{HR}}=-\frac{(1-C_{1})}{2}ln\frac{(1-C_{1})}{2}-\frac{(1+C_{1})}{%
2}ln\frac{(1+C_{1})}{2}.  \label{J_<HR>}
\end{equation}
This quantity is a measure of the inaccuracy introduced in the
observation of observable $\{| e\rangle \langle e|
,| g\rangle \langle g| \} $ by the insertion of cavity $C$.

From the point of view of complementarity the experimental setup of the
Haroche-Ramsey experiment is particularly interesting when also the
information is exploited that is stored in the cavity $C$ field, because
this may add `which way' information to the (non-ideal) interference
information obtained from the measurement of \ the final state of the atom.
This will be discussed in the next sections.

\subsection{The Davidovich-Haroche experiment\label{DavHar}}

By Davidovich et al. \cite{DaHa96} a variation of the Haroche-Ramsey
experiment has been proposed in which a second atom traverses the system some
time after the first one has passed. In \cite{DaHa96} the reason for sending
this second atom is to probe a possible decoherence of the field in cavity $%
C $. We shall discuss this aspect of the experiment in section
\ref{Decoherence}. Here we are interested in the possibility to
consider the 
second atom as yielding information on the cavity $C$ field that might be
useful for determining the path of atom $1.$ We shall demonstrate that the
joint measurement of standard observables $\{ \left| e_{1}\right\rangle
\left\langle e_{1}\right| ,\left| g_{1}\right\rangle \left\langle
g_{1}\right| \} $ and $\{ \left| e_{2}\right\rangle \left\langle
e_{2}\right| ,\left| g_{2}\right\rangle \left\langle g_{2}\right| \} $
in the final state of the atoms can be interpreted as a measurement of a
POVM on the incoming state of atom $1$. We shall neglect decoherence here by
taking a negligible time interval between the atoms. We also restrict
here to the case $\nu =\delta=0$, for which
$E+F=0$, and, hence, (\ref{interf_obs}) is yielding $\hat{Q}_{e}=\left|
g\right\rangle \left\langle g\right| ,\hat{Q}_{g}=\left| e\right\rangle
\left\langle e\right| $. Then, starting with atom $2$ in state $\left|
e_{2}\right\rangle ,$ and using rules (\ref{transformatie
C}) for both atoms, we find for an arbitrary initial state\ $\left| \psi
_{in1}\right\rangle =\alpha \left| e_{1}\right\rangle +\beta \left|
g_{1}\right\rangle $ of atom $1$ the final state 
\[
\left| \Psi _{f}\right\rangle =\frac{1}{4}\left[ 
\begin{array}{c}
\left| e_{1}e_{2}\right\rangle e^{-2i\omega T}\left\{ \alpha \left( |
v_{g}^{\prime }\rangle -2\left| \gamma \right\rangle \right) -i\beta
e^{-i\omega t_{1}}| v_{e}^{\prime }\rangle \right\} + \\ 
\left| e_{1}g_{2}\right\rangle e^{i\omega _{0}\tau }\left\{ -i\alpha
e^{i\omega t_{1}}\left| v_{e}^{\prime }\right\rangle -\beta \left( |
v_{g}^{\prime }\rangle +2\left| \gamma \right\rangle \right) \right\} +
\\ 
\left| g_{1}e_{2}\right\rangle e^{i\omega _{0}\tau }\left\{ -i\alpha
e^{i\omega t_{1}}\left| v_{e}^{\prime }\right\rangle -\beta \left( |
v_{g}^{\prime }\rangle -2\left| \gamma \right\rangle \right) \right\} +
\\ 
\left| g_{1}g_{2}\right\rangle e^{2i\omega T+2i\omega _{0}\tau +i\omega
t_{1}}\left\{ -\alpha e^{i\omega t_{1}}\left( | v_{g}^{\prime
}\rangle +2\left| \gamma \right\rangle \right) +i\beta \left|
v_{e}^{\prime }\right\rangle \right\}
\end{array}
\right] , 
\]
with $\left| v_{e}^{\prime }\right\rangle =| \gamma e^{i2\Phi
}\rangle -| \gamma e^{-i2\Phi }\rangle $ , $|
v_{g}^{\prime }\rangle =| \gamma e^{i2\Phi }\rangle +|
\gamma e^{-i2\Phi }\rangle .$

By putting $\langle \Psi _{f}|\left| e_{1}\right\rangle \left\langle
e_{1}\right| \otimes \left| e_{2}\right\rangle \left\langle e_{2}\right|
\left| \Psi _{f}\right\rangle =$\ $\langle \psi _{in1}|\hat{M}_{e_{1}e_{2}}$\ $%
\left| \psi _{in1}\right\rangle ,$ etc., the POVM of the Davidovich-Haroche
experiment, interpreted as a measurement on atom $1,$ is straightforwardly
found according to 
\begin{equation}
\begin{array}{c}
\hat{M}_{e_{1}e_{2}}=\frac{1}{16}\left( 
\begin{array}{cc}
\parallel | v_{g}^{\prime }\rangle -2\left| \gamma \right\rangle
\parallel ^{2} & -ie^{-i\omega t_{1}}(\langle v_{g}^{\prime }|v_{e}^{\prime
}\rangle -2\langle \gamma |v_{e}^{\prime }\rangle ) \\ 
ie^{i\omega t_{1}}(\langle v_{e}^{\prime }|v_{g}^{\prime }\rangle -2\langle
v_{e}^{\prime }|\gamma \rangle ) & \parallel \left| v_{e}^{\prime
}\right\rangle \parallel ^{2}
\end{array}
\right), \\ 
\hat{M}_{e_{1}g_{2}}=\frac{1}{16}\left( 
\begin{array}{cc}
\parallel \left| v_{e}^{\prime }\right\rangle \parallel ^{2} & -ie^{-i\omega
t_{1}}(\langle v_{e}^{\prime }|v_{g}^{\prime }\rangle +2\langle
v_{e}^{\prime }|\gamma \rangle ) \\ 
ie^{i\omega t_{1}}(\langle v_{g}^{\prime }|v_{e}^{\prime }\rangle +2\langle
\gamma |v_{e}^{\prime }\rangle ) & \parallel | v_{g}^{\prime
}\rangle +2\left| \gamma \right\rangle \parallel ^{2}
\end{array}
\right), \\ 
\hat{M}_{g_{1}e_{2}}=\frac{1}{16}\left( 
\begin{array}{cc}
\parallel \left| v_{e}^{\prime }\right\rangle \parallel ^{2} & -ie^{-i\omega
t_{1}}(\langle v_{e}^{\prime }|v_{g}^{\prime }\rangle -2\langle
v_{e}^{\prime }|\gamma \rangle ) \\ 
ie^{i\omega t_{1}}(\langle v_{g}^{\prime }|v_{e}^{\prime }\rangle -2\langle
\gamma |v_{e}^{\prime }\rangle ) & \parallel | v_{g}^{\prime
}\rangle -2\left| \gamma \right\rangle \parallel ^{2}
\end{array}
\right), \\ 
\hat{M}_{g_{1}g_{2}}=\frac{1}{16}\left( 
\begin{array}{cc}
\parallel | v_{g}^{\prime }\rangle +2\left| \gamma \right\rangle
\parallel ^{2} & -ie^{-i\omega t_{1}}(\langle v_{g}^{\prime }|v_{e}^{\prime
}\rangle +2\langle \gamma |v_{e}^{\prime }\rangle ) \\ 
ie^{i\omega t_{1}}(\langle v_{e}^{\prime }|v_{g}^{\prime }\rangle +2\langle
v_{e}^{\prime }|\gamma \rangle ) & \parallel \left| v_{e}^{\prime
}\right\rangle \parallel ^{2}
\end{array}
\right).
\end{array}
\label{POVMDavHar}
\end{equation}
From these expressions it is immediately clear that averaging over the
initial phase $\omega t_{1}$ makes also the Davidovich-Haroche experiment a
non-ideal measurement of observable $\{ | e\rangle \langle
e| ,| g\rangle \langle g| \} $, the
non-ideality measure (\ref{rowent}) being given by{\normalsize \ } 
\[
J^{\overline{DH}}= 
\begin{array}{c}
-(\frac{1-C_{1}}{2}-\frac{1-C_{1}^{\prime
}}{8})ln(1-\frac{(1-C_{1}^{\prime }) 
}{4(1-C_{1})})-\frac{1-C_{1}^{\prime }}{8}ln\frac{1-C_{1}^{\prime }}{%
4(1-C_{1})}+ \\ 
-(\frac{1+C_{1}}{2}-\frac{1-C_{1}^{\prime
}}{8})ln(1-\frac{(1-C_{1}^{\prime }) 
}{4(1+C_{1})})-\frac{1-C_{1}^{\prime }}{8}ln\frac{1-C_{1}^{\prime }}{%
4(1+C_{1})},
\end{array}
\]
with 
\[
C_{1}^{\prime }+iC_{2}^{\prime }=\langle \gamma e^{i2\Phi }|\gamma
e^{-i2\Phi }\rangle . 
\]
Comparing $J^{\overline{DH}}$ with the corresponding non-ideality
measure (\ref{J_<HR>}) of the Haroche-Ramsey experiment, we find (cf.
figure~\ref{fig2}) that $J^{\overline{DH}}<J^{\overline{HR}}$ for $\gamma 
\neq 0.$ Hence, by taking into account the extra information from the
measurement of the cavity $C$ field the quality of the non-ideal measurement
of the interference observable $\{\hat{Q}_{e},\hat{Q}_{g}\}$ has been
increased. 

In the phase-averaged case the subspace ${\cal H}_{\{
\hat{M}_{m}\} }$ 
of Hilbert-Schmidt space spanned by the operators of POVM (\ref
{POVMDavHar}) is a two-dimensional one. From an informational point of view
the Davidovich-Haroche experiment will be more interesting if it is possible
to avoid the necessity of phase averaging, because in that case ${\cal H}%
_{\{ \hat{M}_{m}\} }$ is three-dimensional. Although, due to the
equality $\hat{M}_{e_{1}e_{2}}+\hat{M}_{g_{1}e_{2}}=(\parallel |
v_{g}^{\prime 
}\rangle -2\left| \gamma \right\rangle \parallel ^{2}+\parallel \left|
v_{e}^{\prime }\right\rangle \parallel ^{2})/16\ast \hat{I},$ the operators of the
POVM are linearly dependent, and, hence, the measurement is not a complete
one, it nevertheless is a generalized measurement, being interpretable,
in the sense defined in sect.~\ref{jointnonid}, as a
joint non-ideal measurement of two incompatible observables. In order to see
this the operators must be ordered in a bivariate way. Due to the
uninformativeness of the marginal $\{
\hat{M}_{e_{1}e_{2}}+\hat{M}_{g_{1}e_{2}},\hat{M}_{e_{1}g_{2}}+
\hat{M}_{g_{1}g_{2}}\}$ the 
only interesting way to do so is according to 
\[
\hat{R}_{mn}=\left( 
\begin{array}{cc}
\hat{M}_{e_{1}e_{2}} & \hat{M}_{e_{1}g_{2}} \\ 
\hat{M}_{g_{1}g_{2}} & \hat{M}_{g_{1}e_{2}}
\end{array}
\right), 
\]
yielding marginals $\{\Sigma _{n}\hat{R}_{mn}\}$ and $\{\Sigma
_{m}\hat{R}_{mn}\}$ 
with

$\hat{M}_{e_{1}e_{2}}+\hat{M}_{e_{1}g_{2}}=\frac{1}{2}\left( 
\begin{array}{cc}
1-C_{1} & C_{2} \\ 
C_{2} & 1+C_{1}
\end{array}
\right) $ and $\hat{M}_{e_{1}e_{2}}+\hat{M}_{g_{1}g_{2}}=\frac{1}{4}\left( 
\begin{array}{cc}
3+C_{1}^{\prime } & -C_{2}^{\prime } \\ 
-C_{2}^{\prime } & 1-C_{1}^{\prime }
\end{array}
\right).$ Here $t_{1\text{ }}$ is taken to be zero. Evidently, both
marginals are depending on the parameter $\Phi $ governing the
distinguishability of the field states.
In agreement with (\ref{2}), for $\gamma \neq 0$ these marginals can be
interpreted as describing non-ideal measurements of two incompatible PVMs of
atom $1$, with non-ideality matrices given by

\begin{eqnarray*}
\left( \lambda \right) &=&\left( 
\begin{array}{cc}
\lambda & 1-\lambda \\ 
1-\lambda & \lambda
\end{array}
\right) ,\left( \mu \right) =\left( 
\begin{array}{cc}
\mu & 1-\mu \\ 
1-\mu & \mu
\end{array}
\right) , \\
\lambda &=&\frac{1}{2}(1+\sqrt{C_{1}^{2}+C_{2}^{2}}),\mu =\frac{1}{2}(1+%
\frac{1}{2}\sqrt{(C_{1}^{\prime }+1)^{2}+C_{2}^{\prime 2}}),
\end{eqnarray*}
yielding non-ideality measures (\ref{rowent}) as

\begin{eqnarray*}
J_{(\lambda )} &=&-\{\lambda \ln (\lambda )+(1-\lambda )\ln (1-\lambda )\},
\\
J_{(\mu )} &=&-\{\mu \ln (\mu )+(1-\mu )\ln (1-\mu )\}.
\end{eqnarray*}
For the parameters $\lambda $ and $\mu $ we find 
\begin{eqnarray*}
\lambda &=&\frac{1}{2}(1+e^{-2\gamma ^{2}\sin ^{2}\Phi }), \\
\mu &=&\frac{1}{2}+\frac{1}{4}[\{1+e^{-2\gamma ^{2}\sin ^{2}2\Phi }\cos
(\gamma ^{2}\sin 4\Phi )\}^{2}+e^{-4\gamma ^{2}\sin ^{2}2\Phi }\sin
^{2}(\gamma ^{2}\sin 4\Phi )]^{1/2}.
\end{eqnarray*}

We shall not bother to calculate the corresponding PVMs, because these 
do not admit a straightforward physical interpretation in terms of the
interference and path observables defined above. From the non-ideality
measures $J_{(\lambda)}$ and $J_{(\mu)}$ it can already be
seen that the two PVMs measured jointly in the Davidovich-Haroche
experiment do not constitute a canonically conjugate pair in the sense
that, if the parameter $\gamma $ is varied, one 
measurement gets more accurate if the other one becomes more non-ideal.
Thus, in both of the limits $\gamma =0$
and $\gamma \rightarrow \infty $ POVM (\ref{POVMDavHar})
is representing a (non-)ideal measurement of the same PVM $\{
\left| e\right\rangle \left\langle e\right| ,\left| g\right\rangle
\left\langle g\right| \} .$
From the plots of $J_{(\lambda )}$ and $J_{(\mu )}$ as functions of
$\gamma $ and $\Phi$ in figure~\ref{fig3} it is also seen that both 
non-ideality measures vanish in the limit $\gamma \rightarrow 0$.
Hence, although the two PVMs measured jointly in this experiment do satisfy
inequality (\ref{3}) for all values of $\gamma$, this does not imply
any complementarity for $\gamma \rightarrow 0$ because the right-hand
side of the inequality is vanishing in this limit due to the fact that the two
PVMs coincide. As will be demonstrated in section \ref{Infasp}, the 
Davidovich-Haroche experiment, for general values of the parameters
$\nu$ and $\delta$, is informationally equivalent to a
measurement in which the second atom is replaced by a measurement of
photon number in the final state of cavity $C$. The absence of
information on the phase of the cavity $C$ field explains the
somewhat non-complementary behavior observed here. In section
\ref{joint} an alternative measurement procedure will be discussed,
better satisfying the canonical notion of complementarity, in which it
is proposed to perform a measurement of the cavity $C$ field also yielding
phase information. 

\section{Informational aspects of the Davidovich-Haroche experiment}
\label{Infasp}
\subsection{Decoherence}
\label{Decoherence}
\noindent The Haroche-Ramsey experiment \cite{BrHa96} was devised in the
first place to probe decoherence in cavity $C$ following the passage of a Rb
atom, entering in state $|e\rangle$. Hence $\beta =0$ in the final state
(\ref{final state}). Restricting to $\nu= \delta=0$ it 
follows from (\ref{final state}) that, conditional on measurement
result $e$ or $g,$ the cavity $C$ field is described by a superposition of
coherent states $( e\rightarrow | v_{e}\rangle =|
\gamma e^{i\Phi }\rangle -| \gamma e^{-i\Phi }\rangle
,g\rightarrow \left| v_{g}\right\rangle =| \gamma e^{i\Phi
}\rangle +| \gamma e^{-i\Phi }\rangle ) $. These
states can be considered as Schr\"{o}dinger cat states if $\gamma $ is
sufficiently large. In \cite{DaHa96} it was proposed to probe, by
sending after a time $T$ a second Rb atom through the system (like the first
atom starting in state $\left| e\right\rangle $), whether
a process of decoherence is active by which these
superpositions could decay to a mixture of the coherent states $| \gamma
e^{i\Phi }\rangle $ and $| \gamma e^{-i\Phi }\rangle $. In
\cite{BrHa96} it was 
concluded from the observed two-atom correlations\ that a decoherence effect
obtains. In particular it was inferred from the experimental data that the
decoherence time is much shorter than the decay time that can be attributed
to loss of photons from cavity $C$. In the present section this latter
conclusion is challenged. It is demonstrated that, in agreement with a
result obtained by Vitali et al. \cite{ViToMi98}, the second Rb atom can only
yield information on cavity $C$'s photon number. Hence, any change of the
measurement results obtained for this atom should be attributed to a change
of photon number. Of course, this does not imply the absence of decoherence
due to decay of pure phase correlations. However, this measurement is not
sensitive to it.

In order to demonstrate this we have to determine which information is
obtained on the cavity $C$ field by the measurement of the second atom. The
corresponding POVM can be found by equating, for an arbitrary initial coherent
state $\left| \alpha \right\rangle $ of the cavity field, the final state
probabilities $p_{e}$ and $p_{g}$ to expectation values $\langle \alpha
| \hat{M}_{e}| \alpha \rangle $ and $\langle
\alpha | 
\hat{M}_{g}$ $| \alpha \rangle $, respectively. With $| \Psi
_{f}\rangle =\frac{1}{2}| e_{2}\rangle ( E|
\alpha e^{i\Phi }\rangle +F| \alpha e^{-i\Phi }\rangle
) +\frac{1}{2}| g_{2}\rangle ( J| \alpha e^{i\Phi
}\rangle +K| \alpha e^{-i\Phi }\rangle ) $ we find, once again
restricting to $\nu=\delta =0$, 
\begin{eqnarray*}
p_{e} &=& \langle \Psi _{f}| e_{2}\rangle
\langle e_{2}| \Psi _{f}\rangle  =\frac{1}{4}%
\parallel | \alpha e^{i\Phi }\rangle -| \alpha e^{-i\Phi
}\rangle \parallel ^{2}=\frac{1}{4}\langle \alpha |
2-e^{2i\Phi \hat{a}^{\dagger }\hat{a}}-e^{-2i\Phi \hat{a}^{\dagger
}\hat{a}}| \alpha 
\rangle , \\
p_{g} &=& \langle \Psi _{f}| g_{2}\rangle
\langle g_{2}| \Psi _{f}\rangle  =\frac{1}{4}%
\parallel | \alpha e^{i\Phi }\rangle +| \alpha e^{-i\Phi
}\rangle \parallel ^{2}=\frac{1}{4}\langle \alpha |
2+e^{2i\Phi \hat{a}^{\dagger }\hat{a}}+e^{-2i\Phi \hat{a}^{\dagger
}\hat{a}}| \alpha \rangle ,\ 
\end{eqnarray*}
from which we obtain
\[
\hat{M}_{e}=\sin ^{2}\Phi \hat{a}^{\dagger }\hat{a},\;
\hat{M}_{g}=\cos ^{2}\Phi \hat{a}^{\dagger }\hat{a}.
\]
Note that this result holds independent of phase averaging. It is easily seen
that POVM $\{ \hat{M}_{e},\hat{M}_{g}\} $ represents a
non-ideal 
measurement of the number observable $\hat{N}=\hat{a}^{\dagger
}\hat{a}=\sum_{n=0}^{\infty }n| n\rangle \langle
n| $ in the sense of definition (%
\ref{1}): 
\begin{eqnarray*}
\hat{M}_{e} &=&\sum_{n=0}^{\infty }\lambda _{en}\left| n\right\rangle \left\langle
n\right| ,\hat{M}_{g}=\sum_{n=0}^{\infty }\lambda _{gn}\left| n\right\rangle
\left\langle n\right| , \\
\lambda _{en} &=&\sin ^{2}\Phi n,\lambda _{gn}=\cos ^{2}\Phi n.
\end{eqnarray*}
Using (\ref{biorth}) it is possible to calculate the projection
$\hat{\rho}_{\{\hat{M}_{m}\}}$ of the density operator $\hat{\rho} $ on the
subspace spanned by $%
\hat{M}_{e}$ and $\hat{M}_{g},$ representing the information that is
obtained by a 
measurement of POVM $\{\hat{M}_{e},\hat{M}_{g}\}.$ Due to the
infinite-dimensionality of the Hilbert space of the field this must be done
with some care because the operators are not Hilbert-Schmidt operators then.
For this reason the dimension must be truncated. Restricting to an arbitrary
large but finite value $D,$ we get: 
\begin{eqnarray*}
\hat{M}_{m}^{\prime } &=&\sum_{n=0}^{D}[\beta _{me}\sin ^{2}\Phi
n+\beta _{mg}\cos^{2}\Phi n]\left| n\right\rangle \left\langle n\right|
,m=e,g, \\ 
\beta _{ee} &=&\frac{r-s}{(1+r-s)(D-r-s)-s^{2}},\;\beta _{ge}=\beta_{eg}=
\frac{-s}{(1+r-s)(D-r-s)-s^{2}}, \\
\beta _{gg} &=&\frac{D-r-s}{(1+r-s)(D-r-s)-s^{2}},
\end{eqnarray*}
with $r=\sum_{n=1}^{D}\cos ^{2}\Phi n,s=\sum_{n=1}^{D}\sin ^{2}\Phi n\cos
^{2}\Phi n.$ Then 
\begin{eqnarray}
\hat{\rho}_{\{\hat{M}_{m}\}} &=&\lim_{D\rightarrow \infty }\sum_{n,n^{\prime
}=0}^{D}[(\beta _{ee}\sin ^{2}\Phi n^{\prime }+\beta _{ge}\cos ^{2}\Phi
n^{\prime })\sin ^{2}\Phi n+ \\
&&+(\beta _{eg}\sin ^{2}\Phi n^{\prime }+\beta _{gg}\cos ^{2}\Phi n^{\prime
})\cos ^{2}\Phi n]\left\langle n^{\prime }\right| \hat{\rho} \left| n^{\prime
}\right\rangle \left| n\right\rangle \left\langle n\right| .  \nonumber
\end{eqnarray}
Note that, although $\beta _{mm^{\prime }}\rightarrow 0$ if
$D\rightarrow \infty ,$ yet  $\hat{\rho}_{\{\hat{M}_{m}\}}\neq
\hat{O}.$ Thus, it is easily verified that also 
in the limit $D\rightarrow \infty $ $Tr\hat{\rho}_{\{\hat{M}_{m}\}}=1,$
and $Tr\hat{\rho}_{\{\hat{M}_{m}\}}\hat{M}_{m}=Tr\hat{\rho}
\hat{M}_{m},m=e,g,$ the latter equality explicitly demonstrating 
that $\hat{\rho}_{\{\hat{M}_{m}\}}$ contains the same
information on the 
measurement results of POVM $\{\hat{M}_{e},\hat{M}_{g}\}$ as does
$\hat{\rho} .$ 

Although a measurement of this POVM can distinguish between a mixture of the
states $\left| v_{e}\right\rangle $ and $\left| v_{g}\right\rangle $ and a
mixture of the states $| \gamma e^{i\Phi }\rangle $ and $|
\gamma e^{-i\Phi }\rangle ,$ this is only so because the probability
distributions of the photon {\em number} observable are different in the two
mixtures. Decoherence, not accompanied by a change of photon number, cannot
be observed using the Davidovich-Haroche experiment.

\subsection{Informational equivalence of second atom and number
measurement}
\label{equiv}
In this section it will be demonstrated that the Davidovich-Haroche
experiment, in which a second atom is used as a probe of the cavity $C$
field, is informationally equivalent to a Haroche-Ramsey experiment
in which the `which-way'
information is obtained by measuring photon number. This will be done by
considering the informational aspects of the measurement, introduced in
section \ref{genmeasinfo}. From the informational point of view the
important feature is the structure of the subspaces ${\cal H}_{\{
\hat{M}_{m}\} }$ of Hilbert-Schmidt space, spanned by the operators
$\hat{M}_{m}$ generating the POVM, as a function of the experimental
parameters. 
In order to be completely general, in this section we allow the different
parameters to take arbitrary values. We first determine the POVM of the
Haroche-Ramsey experiment in which cavity $C$ photon number is measured in
coincidence with a determination of the final state of the atom. This POVM
is found from (\ref{final state}) by the equalities

\[
\begin{array}{c}
p_{en}=\langle \Psi_{f} | e\rangle \langle e| \otimes | n\rangle
\langle n| \Psi_{f}\rangle =\langle \psi_{in}| \hat{M}_{en}|
\psi_{in}\rangle, \\ 
p_{gn}=\langle \Psi _{f} | g\rangle \langle g| \otimes | n\rangle 
\langle n| \Psi_{f}\rangle =\langle \psi_{in}| \hat{M}_{gn}| 
\psi_{in}\rangle,
\end{array}
\]
in which $p_{en}$ and $p_{gn}$ are the measured joint probabilities, and $%
\{| n\rangle \langle n| \}$ is the PVM corresponding
to the spectral measure of the photon number observable. With $t_1=0$
the operators $\hat{M}_{en}$ and $\hat{M}_{gn}$ are found as
\begin{eqnarray*}
\hat{M}_{en} &=&\frac{e^{-\gamma ^{2}}\gamma ^{2n}}{n!}\left( 
\begin{array}{cc}
\sin^2\frac{\varphi _{n}}{2}
+\delta^{2}\cos ^{2}\frac{\varphi _{n}}{2} 
& -e^{-i\psi }\frac{\sqrt{1-\delta^2}}{2}[\sin \varphi
_{n}+2i\delta\cos ^{2}\frac{%
\varphi _{n}}{2}] \\ 
-e^{i\psi }\frac{\sqrt{1-\delta^2}}{2}[\sin \varphi
_{n}-2i\delta \cos 
^{2}\frac{\varphi _{n}}{2}] & (1-\delta^2)\cos ^{2}\frac{\varphi _{n}}{2%
}
\end{array}
\right), \\
\hat{M}_{gn} &=&\frac{e^{-\gamma ^{2}}\gamma ^{2n}}{n!}\left( 
\begin{array}{cc}
(1-\delta^2) \cos ^{2}\frac{\varphi _{n}}{2} 
& e^{-i\psi }\frac{\sqrt{1-\delta^2}}{2}[\sin \varphi_{n}+2i\delta \cos
^{2}\frac{\varphi _{n}}{2}] \\ 
e^{i\psi }\frac{\sqrt{1-\delta^2}}{2}[\sin \varphi
_{n}-2i\delta\cos ^{2} 
\frac{\varphi _{n}}{2}] & 
\sin^2\frac{\varphi _{n}}{2}
+\delta^{2}\cos ^{2}\frac{\varphi _{n}}{2} 
\end{array}
\right),
\end{eqnarray*}
in which $\varphi _{n}=2n\Phi +\nu \tau +2\psi ,\psi =\arg (S_{1}).$

Excluding $\delta =\pm 1$, for which POVM
$\{\hat{M}_{en},\hat{M}_{gn}\}$ reduces to a trivial refinement of
PVM $\{|e\rangle\langle e|,|g\rangle\langle g|\} $,
for most values of the parameters the 
operators $\hat{M}_{en}$ and $\hat{M}_{gn},n=0,1..$ 
span the whole Hilbert-Schmidt space of operators on a 2-dimensional
Hilbert space. Hence, in general the measurement is a complete measurement.
The parameter values for which the measurement is incomplete can be found by
looking for Hermitian operators $\hat{T}$ that are orthogonal to all
$\hat{M}_{en}$ and $ \hat{M}_{gn}.$ Thus, $Tr \hat{M}_{en}\hat{T}= Tr
\hat{M}_{gn}\hat{T}=0,n=0,..$ We find 

$\Phi =\frac{\pi }{2},\delta\neq 0:$

$\hat{T}=\left( 
\begin{array}{cc}
-\sqrt{1-\delta^2}\tan (\nu \tau +2\psi ) & e^{-i\psi
}[1-i\delta \tan (\nu \tau +2\psi )] \\ 
e^{i\psi }[1+i\delta \tan (\nu \tau +2\psi )] & 
\sqrt{1-\delta^2}\tan (\nu \tau +2\psi )
\end{array}
\right) ;$

$\Phi \neq \frac{\pi }{2},\delta=0:$

$\hat{T}=\left( 
\begin{array}{cc}
0 & ie^{-i\psi } \\ 
-ie^{i\psi } & 0
\end{array}
\right) ;$

$\Phi =\frac{\pi }{2},\delta=0:$

$\hat{T}=\left( 
\begin{array}{cc}
\tan (\nu \tau +2\psi ) & e^{-i\psi } \\ 
e^{i\psi } & \tan (\nu \tau +2\psi )
\end{array}
\right) ,\left( 
\begin{array}{cc}
0 & ie^{-i\psi } \\ 
-ie^{i\psi } & 0
\end{array}
\right) .$
Barring $\delta=\pm 1$, for all other values of the parameters no
solution for $\hat{T}$ can be 
found. Hence, specializing the parameters of the experiment to either
$\Phi =\frac{%
\pi }{2}$, or to the $\frac{\pi }{2}$ pulse condition $\delta=0$, reduces
the dimension of the subspace spanned by the operators of the POVM 
to 3, the dimensionality being further reduced if both conditions are
simultaneously satisfied. By determining in the same way the
Hilbert-Schmidt operators that are orthogonal to the operators of
POVM (\ref{POVMDavHar}) it is straightforward to prove that the 
Davidovich-Haroche experiment, 
discussed in section \ref{DavHar}, has exactly the same structure of
subspaces, thus demonstrating the informational equivalence of these
experiments for all values of the parameters.

The subspace 
structure is not essentially changed by taking the detuning parameter $\nu
=0.$ Since then also $\psi =0$ the operators $\hat{T}$, found above, are
particularly simple, viz. $%
 \left( 
\begin{array}{cc}
0 & i \\ 
-i & 0
\end{array}
\right) $ and $\left( 
\begin{array}{cc}
0 & 1 \\ 
1 & 0
\end{array}
\right) .$ These are two orthogonal vectors, constituting together with the
operators $\left( 
\begin{array}{cc}
1 & 0 \\ 
0 & 0
\end{array}
\right) $ and $\left( 
\begin{array}{cc}
0 & 0 \\ 
0 & 1
\end{array}
\right) $ an orthogonal basis of Hilbert-Schmidt space. Due to the
uniqueness of the Hermitian projection operator ${\cal P}_{\{
\hat{M}_{m}\} }$ this makes it particularly easy to calculate the
projected density operator ${\cal P}_{\{ \hat{M}_{m}\} }\hat{\rho}
$ representing the 
information about the density operator $\hat{\rho} $ provided by the
measurement. We find (with $\nu =0):$

$\Phi =\frac{\pi }{2},\delta\neq 0:{\cal P}_{\{
\hat{M}_{m}\} }\hat{\rho} =\left( 
\begin{array}{cc}
\rho _{11} & \frac{\rho _{12}-\rho _{21}}{2} \\ 
-\frac{\rho _{12}-\rho _{21}}{2} & \rho _{22}
\end{array}
\right) ;$

$\Phi \neq \frac{\pi }{2},\delta = 0:{\cal P}_{\{
\hat{M}_{m}\} }\hat{\rho} =\left( 
\begin{array}{cc}
\rho _{11} & \frac{\rho _{12}+\rho _{21}}{2} \\ 
\frac{\rho _{12}+\rho _{21}}{2} & \rho _{22}
\end{array}
\right) ;$

$\Phi =\frac{\pi }{2},\delta = 0:{\cal P}_{\{
\hat{M}_{m}\} }\hat{\rho} =\left( 
\begin{array}{cc}
\rho _{11} & 0 \\ 
0 & \rho _{22}
\end{array}
\right) .$

Note that all ${\cal P}_{\{ \hat{M}_{m}\} }\hat{\rho} $ are
non-negative (cf. Appendix). In the $\Phi =\frac{\pi }{2},\delta=0$ case only information is
obtained on the diagonal elements of $\hat{\rho} $ because for these parameter
values the measurement is a non-ideal measurement of PVM $\{ \left|
e\right\rangle \left\langle e\right| ,\left| g\right\rangle \left\langle
g\right| \} .$ 
It is clear from this that from the informational point of view
our parameter choice in section \ref{DavHar} was not completely appropriate
for the purpose of reconstructing the initial density operator. By
restricting to $\frac{\pi }{2}$ pulses the measurement cannot retrieve $%
\mathop{\rm Im}%
\rho _{12}$. For a complete determination of $\hat{\rho} $ it is
necessary that $ \Phi \neq \frac{\pi }{2}$ and $\delta \neq 0$ (keeping
$\delta \neq\pm 1$).

One remark is in order here. Since the special parameter values $\Phi =\frac{%
\pi }{2}$ and $\delta=0$ constitute sets of measure zero
within the set of all possible values of the parameters, it might be
thought that these special values are physically irrelevant because they
cannot be attained in practice. In a strict sense this is correct. However,
even though in practice the POVMs are complete, this does not mean that the
subspace structure is unimportant. As a matter of fact, if the parameter
values are near the special values given above, then the quantities 
$Tr (\hat{\rho} -\hat{\rho}_{\{\hat{M}_{m}(\Phi =\frac{\pi
}{2},\delta=0)\}})\hat{M}_{m}$ will be very small. This means that the experimental
error in the determination of these quantities is relatively large.
Hence, the experimental probabilities will yield relatively poor 
information about the components of the Hilbert-Schmidt vector $\hat{\rho} $
orthogonal to ${\cal H}_{\{\hat{M}_{m}(\Phi =\frac{\pi
}{2},\delta=0)\}}$. Stated 
differently, sets $\{\hat{M}_{m}\}$ constituting bases of 
Hilbert-Schmidt space need not be equivalent from an informational point of
view, the quality of the information being largest for an orthogonal basis.
A measure of this quality will be discussed elsewhere. 

\section{The Haroche-Ramsey experiment as a joint non-ideal measurement of
interference and path observables}
\label{joint}

Rather than exploiting a second atom, or, equivalently, measuring
observable $\{ \sin ^{2}\Phi \hat{a}^{\dagger }\hat{a},\cos
^{2}\Phi \hat{a}^{\dagger }\hat{a}\} $ of the cavity $C$ field
jointly with observable $%
\{ \left| e\right\rangle \left\langle e\right| ,\left| g\right\rangle
\left\langle g\right| \} ,$ 
we consider here the field observable $\{
\hat{M}_{+},\hat{M}_{-}\} $ defined by 
\[
\hat{M}_{+}=\frac{1}{\pi }%
\mathrel{\mathop{\int }\limits_{{C}^{+}}}%
d^{2}\alpha \left| \alpha \right\rangle \left\langle \alpha \right| \text{ ; 
}\hat{M}_{-}=\frac{1}{\pi }%
\mathrel{\mathop{\int }\limits_{{C}^{-}}}%
d^{2}\alpha \left| \alpha \right\rangle \left\langle \alpha \right| ,
\]
$\left| \alpha \right\rangle $ a coherent state, and the integrations being
over the upper $({C}^{+})$\ and lower $({C}^{-})$
complex half-planes, respectively. This observable is a coarsening of the
observable $\{\frac{1}{\pi }\left| \alpha \right\rangle \left\langle \alpha
\right| \}$ measured in the eight-port homodyning detection method
\cite{YuSha80,Qfunct}. For $\gamma \rightarrow \infty $ 
we have $\left\langle \gamma
e^{i\Phi }\right| \hat{M}_{m}\left| \gamma e^{i\Phi }\right\rangle \rightarrow
\delta _{m+},\left\langle \gamma e^{-i\Phi }\right| \hat{M}_{m}\left| \gamma
e^{-i\Phi }\right\rangle \rightarrow \delta _{m-}$. In this limit
POVM\ $\{ \hat{M}_{+},\hat{M}_{-}\} $\ is evidently yielding 
information on the phase shift $\Phi $ caused by the Rb atom, and, hence,
is providing path information. In the Haroche-Ramsey experiment \cite{BrHa96}
the value of $\gamma $ was finite ($\gamma \approx 3),$ 
causing the distinguishability of the states to be only partial. As
will be seen in the following, this loss of path information is
compensated by the interference information obtained by measuring a
quantity of the $C$ field yielding information on both number and phase.

In order to interpret the experiment as a measurement in the initial state $%
\left| \psi _{in}\right\rangle $ we put 
\[
p_{m\pm }=\left\langle \Psi _{f}\right| \left( \left| m\right\rangle
\left\langle m\right| \otimes \hat{M}_{\pm }\right) \left| \Psi
_{f}\right\rangle 
=\left\langle \psi _{in}\right| \hat{M}_{m\pm }\left| \psi _{in}\right\rangle
,\;m=e,g,\]
yielding a POVM $\{
\hat{M}_{e+},\hat{M}_{e-},\hat{M}_{g+},\hat{M}_{g-}\} $ with
elements given by 
\begin{equation}
\begin{array}{c}
\hat{M}_{e\pm }=\frac{1}{2}[(1\pm A-C_{1})\left| S_{1}\right|
^{2}\hat{P}_{+}+(1\mp 
A-C_{1})\left| S_{2}\right| ^{2}\hat{P}_{-}+C_{1}\hat{Q}_{e}+C_{2}\hat{S}], \\ 
\hat{M}_{g\pm }=\frac{1}{2}[1\mp A\delta]\hat{I}-\hat{M}_{e\mp },\;\delta=|
S_{1}|^{2}-| S_{2}|^{2}.
\end{array}
\label{POVMjoint}
\end{equation}
Here $\{ \hat{P}_{+},\hat{P}_{-}\} $ and $\{
\hat{Q}_{e},\hat{Q}_{g}\} $ are 
the path and interference observables (\ref{path_obs}) and (\ref{interf_obs}%
) defined above. The constant $A$ is given by 
\[
A=%
\mathop{\rm erf}%
(\gamma \sin \Phi ),
\]
and $C_{1}$ and $C_{2}$ are given by (\ref{C}). The operator $\hat{S}$
is defined according to 
\[
\hat{S}=ie^{-i\nu \tau }S_{1}^{\ast }S_{2}|p_{+}\rangle \langle
p_{-}|\text{-}%
ie^{i\nu \tau }S_{1}^{\ast }S_{2}|p_{-}\rangle \langle p_{+}|,
\]
$|p_{\pm }\rangle $ being given by (\ref{path_obs}).

In the phase averaged case the experiment described
by this POVM once again is a non-ideal measurement of PVM $\{
\left| e\right\rangle \left\langle e\right| ,\left| g\right\rangle
\left\langle g\right| \} .$ Restricting to $\nu =\delta=0$ we find

\[
\overline{\hat{M}_{e\pm }}=\frac{1}{4}(1-C_{1})\left| e\right\rangle
\left\langle 
e\right| +%
\frac{1}{4}(1+C_{1})\left| g\right\rangle \left\langle g\right|
,\overline{\hat{M}_{g\pm }}=%
\frac{1}{4}(1+C_{1})\left| e\right\rangle \left\langle e\right| +\frac{1}{4}%
(1-C_{1})\left| g\right\rangle \left\langle g\right| , \]
yielding for the non-ideality measure the same outcome (\ref{J_<HR>})
as obtained in
the experiment in which no measurement is performed on the cavity $C$ field.
Evidently, in the phase averaged case such a measurement does not improve the
information. Indeed, the two measurements are equivalent in the sense
defined in \cite{MadM90}.

We shall now consider POVM (\ref{POVMjoint}) when no phase
averaging is performed. We should exclude $\delta =\pm 1$ also here 
because then also POVM $\{\hat{M}_{e\pm},\hat{M}_{g\pm}\}$ reduces to a
trivial refinement of PVM $\{|e\rangle\langle e|,|g\rangle\langle g|\}
$ (this actually holds true for any choice of the observable of the
cavity $C$ field). In the limit $\gamma =0$
the POVM\ reduces to $\{ 
\frac{1}{2}\hat{Q}_{e},%
\frac{1}{2}\hat{Q}_{e},\frac{1}{2}\hat{Q}_{g},\frac{1}{2}\hat{Q}_{g}\}
,$ representing a 
trivial refinement of the interference observable $\{
\hat{Q}_{e},\hat{Q}_{g}\} $. On the other hand, for $\gamma
\rightarrow \infty $ 
the POVM reduces to the trivial refinement $\{ \left| S_{1}\right|
^{2}\hat{P}_{+},\left| S_{2}\right| ^{2}\hat{P}_{-},\left| S_{2}\right|
^{2}\hat{P}_{+},\left| 
S_{1}\right| ^{2}\hat{P}_{-}\} $ of the path observable $\{
\hat{P}_{+},\hat{P}_{-}\} $. We shall now demonstrate that if the
$\frac{\pi }{2}$ pulse condition $\delta=0$ is satisfied ($\nu $
arbitrary) and $\Phi =\frac{\pi }{2},$ the Haroche-Ramsey experiment can be
interpreted, in the sense of section \ref{jointnonid}, as a joint non-ideal
measurement of $\ $the incompatible observables
$\{\hat{Q}_{e},\hat{Q}_{g}\}$ and $%
\{ \hat{P}_{+},\hat{P}_{-}\} $. To see this we define the
bivariate POVM $ \{ \hat{M}_{mn}\} $:

\begin{equation}
\{ \hat{M}_{mn}\} =\left( 
\begin{array}{cc}
\hat{M}_{e+} & \hat{M}_{g+} \\ 
\hat{M}_{e-} & \hat{M}_{g-}
\end{array}
\right).  \label{definition of Mmn}
\end{equation}
For the two marginals we find, respectively, 
\begin{equation}
\left( 
\begin{array}{l}
\hat{M}_{e+}+\hat{M}_{g+} \\ 
\hat{M}_{e-}+\hat{M}_{g-}
\end{array}
\right) =\stackrel{\left( \lambda _{mn}\right) }{\overbrace{\frac{1}{2}\left(
\begin{array}{ll}
1+%
\mathop{\rm erf}%
(\gamma ) & 1-%
\mathop{\rm erf}%
(\gamma ) \\ 
1-%
\mathop{\rm erf}%
(\gamma ) & 1+%
\mathop{\rm erf}%
(\gamma )
\end{array}
\right) }}\left( 
\begin{array}{l}
\hat{P}_{+} \\ 
\hat{P}_{-}
\end{array}
\right)   \label{horizontale_marg}
\end{equation}
and 
\begin{equation}
\left( 
\begin{array}{l}
\hat{M}_{e+}+\hat{M}_{e-} \\ 
\hat{M}_{g+}+\hat{M}_{g-}
\end{array}
\right) =\stackrel{\left( \mu _{mn}\right) }{\overbrace{\frac{1}{2}\left( 
\begin{array}{ll}
1+e^{-2\gamma ^{2}} & 1-e^{-2\gamma ^{2}} \\ 
1-e^{-2\gamma ^{2}} & 1+e^{-2\gamma ^{2}}
\end{array}
\right) }}\left( 
\begin{array}{l}
\hat{Q}_{e} \\ 
\hat{Q}_{g}
\end{array}
\right) \text{.}  \label{verticale marg}
\end{equation}
In the limits $\gamma \rightarrow \infty $ and $\gamma =0$ the marginals
describe ideal measurements of path and interference, respectively. The
non-ideality measures $J_{(\lambda )}$ and $J_{(\mu )}$ corresponding to the
non-ideality matrices $\left( \lambda _{mn}\right) $ and $\left( \mu
_{mn}\right) $ are found according to 
\[
\begin{array}{c}
J_{(\lambda )}=-\frac{1}{2}\left[ \left( 1+%
\mathop{\rm erf}%
(\gamma )\right) \ln \left( \frac{1+%
\mathop{\rm erf}%
(\gamma )}{2}\right) +\left( 1-%
\mathop{\rm erf}%
(\gamma )\right) \ln \left( \frac{1-%
\mathop{\rm erf}%
(\gamma )}{2}\right) \right] \text{,} \\ 
\text{ } \\ 
J_{(\mu )}=-\frac{1}{2}\left[ \left( 1+e^{-2\gamma ^{2}}\right) \ln \left( 
\frac{1+e^{-2\gamma ^{2}}}{2}\right) +\left( 1-e^{-2\gamma ^{2}}\right) \ln
\left( \frac{1-e^{-2\gamma ^{2}}}{2}\right) \right] \text{.}
\end{array}
\]
For the special values of the parameters considered here we have $|\langle
p_{i}|q_{j}\rangle |=1/\sqrt{2},i=\pm ,j=e,g.$ Hence, for the observables $%
\{\hat{Q}_{e},\hat{Q}_{g}\}$ and $\{
\hat{P}_{+},\hat{P}_{-}\} $ the right-hand side of 
inequality (\ref{3}) is non-vanishing, and it is impossible that both $%
J_{(\lambda )}$ and $J_{(\mu )}$ are equal to zero. In figure~\ref{fig4} $%
J_{(\lambda )}$ is plotted versus $J_{(\mu )}$ as a function of the
parameter $\gamma .$ The resulting curve clearly exhibits the idea of
complementarity expressed by inequality (\ref{3}): the experiment
constitutes a less accurate measurement of the interference observable
as the path observable is determined more accurately by increasing
$\gamma$ (and vice versa). It is impossible that both $J_{(\lambda )}$
and $J_{(\mu )}$ simultaneously have small values.

The nice feature of the $\Phi =\frac{\pi }{2}$ condition is that the $\gamma 
$ dependence of the marginals is completely taken into account by the
non-ideality matrices $\left( \lambda _{mn}\right) $ and $\left( \mu
_{mn}\right) ,$ PVMs $\{\hat{Q}_{e},\hat{Q}_{g}\}$ and $\{
\hat{P}_{+},\hat{P}_{-}\} $ 
being independent of $\gamma .$ This feature is partly lost if we allow
values  $\Phi \neq \frac{\pi }{2}$. For general values of the parameters we
can represent POVM (\ref{POVMjoint}) in the following way:

\begin{eqnarray*}
\hat{M}_{e+} &=&\frac{1}{2}[(1+A)|S_{1}|^{2}|p_{+}\rangle \langle
p_{+}|+(1-A)|S_{2}|^{2}|p_{-}\rangle \langle p_{-}|+\{Ce^{-i\nu \tau
}S_{1}^{\ast }S_{2}\}|p_{+}\rangle \langle p_{-}|+h.c.\}] \\
\hat{M}_{e-} &=&\frac{1}{2}[(1-A)|S_{1}|^{2}|p_{+}\rangle \langle
p_{+}|+(1+A)|S_{2}|^{2}|p_{-}\rangle \langle p_{-}|+\{Ce^{-i\nu \tau
}S_{1}^{\ast }S_{2}\}|p_{+}\rangle \langle p_{-}|+h.c.\}] \\
\hat{M}_{g+} &=&\frac{1}{2}[(1+A)|S_{2}|^{2}|p_{+}\rangle \langle
p_{+}|+(1-A)|S_{1}|^{2}|p_{-}\rangle \langle p_{-}|-\{Ce^{-i\nu \tau
}S_{1}^{\ast }S_{2}\}|p_{+}\rangle \langle p_{-}|+h.c.\}] \\
\hat{M}_{g-} &=&\frac{1}{2}[(1-A)|S_{2}|^{2}|p_{+}\rangle \langle
p_{+}|+(1+A)|S_{1}|^{2}|p_{-}\rangle \langle p_{-}|-\{Ce^{-i\nu \tau
}S_{1}^{\ast }S_{2}\}|p_{+}\rangle \langle p_{-}|+h.c.\}].
\end{eqnarray*}
From this we find as one marginal

\[
\left( 
\begin{array}{l}
\hat{M}_{e+}+\hat{M}_{g+} \\ 
\hat{M}_{e-}+\hat{M}_{g-}
\end{array}
\right) =\stackrel{\left( \lambda _{mn}^{\prime }\right) }{\overbrace{\frac{1%
}{2}\left( 
\begin{array}{ll}
1+A & 1-A \\ 
1-A & 1+A
\end{array}
\right) }}\left( 
\begin{array}{l}
\hat{P}_{+} \\ 
\hat{P}_{-}
\end{array}
\right) , 
\]
$A=%
\mathop{\rm erf}%
(\gamma \sin \Phi ),$ which still is a non-ideal measurement of the the path
observable $\{\hat{P}_{+},\hat{P}_{-}\}$. However for the other
marginal we get 
\[
\left( 
\begin{array}{l}
\hat{M}_{e+}+\hat{M}_{e-} \\ 
\hat{M}_{g+}+\hat{M}_{g-}
\end{array}
\right) =\stackrel{\left( \mu _{mn}^{\prime }\right) }{\overbrace{\left( 
\begin{array}{cc}
\mu & 1-\mu \\ 
1-\mu & \mu
\end{array}
\right) }}\left( 
\begin{array}{c}
\hat{Q}_{e}^{\prime } \\ 
\hat{Q}_{g}^{\prime }
\end{array}
\right) , 
\]
with $\mu =(1-\sqrt{1-(1-\delta^2)(1-|C|^2)})/2$ and $%
\{\hat{Q}_{e}^{\prime }$,$\hat{Q}_{g}^{\prime }\}$ different from\
$\{\hat{Q}_{e}$,$\hat{Q}_{g}\}.$ 
Since PVM $\{\hat{Q}_{e}^{\prime }$,$\hat{Q}_{g}^{\prime }\}$ turns
out to be 
dependent on $\gamma $ the experiment no longer is a joint measurement of
one stable PVM pair when varying $\gamma .$ Nevertheless for each set of
parameters PVM\ $\{\hat{Q}_{e}^{\prime }$,$\hat{Q}_{g}^{\prime }\}$
is incompatible with the path observable, and inequality (\ref{3}) is
satisfied by the non-ideality measures $J_{(\lambda ^{\prime })}$ and
$J_{(\mu ^{\prime })}$. Since the parameters $A$ and $|C|$ 
depend on $\gamma $ and $\Phi $ only as $\gamma \sin \Phi ,$ this
latter quantity, together with $\delta$, is determining the measure of
complementarity of observables $ \{\hat{Q}_{e}^{\prime
}$,$\hat{Q}_{g}^{\prime }\}$ and $\{\hat{P}_{+},\hat{P}_{-}\}$.
By comparing figures \ref{fig5} and \ref{fig3} it is seen that,
contrary to the 
Davidovich-Haroche experiment, these observables are complementary for both
$\gamma=0$ and $\gamma\rightarrow \infty,$ complementarity being
largest for $\delta=0$.

The measurement represented by POVM (\ref{POVMjoint}) is not a
complete one. It can be verified that, if $\gamma\neq 0$, the operator
$\hat{T}=iCe^{-i\nu\tau}S^*_1S_2|p_+\rangle\langle p_-|+h.c.$ is
orthogonal to all operators of the POVM. However, for $\delta \neq
\pm1$ no parameter values 
exist for which subspace ${\cal H}_{\{\hat{M}_{m}\} }$ has
dimension smaller than $3$. This demonstrates the informational
superiority of the present measurement, based on homodyning, over 
the one measuring photon number. By refining the partition
$({\bf{C^+,C^-}})$ of the complex plane POVM (\ref{POVMjoint}) can
easily be refined to one spanning the whole Hilbert-Schmidt space
of $2\times 2$ matrices, allowing a complete determination of the
incoming state $\psi_{in}$ of the atom. 

\section{Summary and conclusions}

In this paper we studied a number of atomic beam 
experiments related to the Ramsey experiment. Whereas this latter
experiment is a pure interference measurement, in the experiments
studied here also `which-way' information can be obtained. This is achieved
by inserting a third microwave cavity, $C$, between the ones already
present in the Ramsey experiment, and measuring some observable of the
cavity $C$ field after the atom has passed. Three different
measurement arrangements were considered. In the first (referred to as the
Davidovich-Haroche experiment) a second atom was used as a probe,
in the second a measurement of photon number is performed instead. In
the third arrangement homodyne optical detection of the cavity $C$
field is contemplated. The experiments were demonstrated to yield new
examples of generalized measurements, to be described by positive
operator-valued measures. POVMs were calculated explicitly for
different values of the experimental parameters. 

The experiments are
interesting for two reasons. In the first place they can be interpreted
as joint non-ideal measurements of incompatible observables, thus
clarifying the notion of complementarity. It was found that, although
all measurements satisfy inequality (\ref{3}), only the last experiment
exhibits complementarity in the sense that there exist two limiting
values of the experimental parameters, for one of which the measurement is
a pure interference measurement in which `which-way' information is
maximally disturbed, whereas in the other limit it is a pure
`which-way' measurement in which no interference can be observed. It was
demonstrated that in general by the other measurement arrangements this
``classical'' type of complementarity need not be satisfied.
In the second place, comparison of the different measurements can give
insight into the question of which information is provided by a
(generalized) quantum mechanical measurement. For this purpose the
subspaces of Hilbert-Schmidt space, spanned by the operators of the
POVM, were determined for different measurement arrangements and
different values of the parameters. It was found that a measurement of
the second atom in the Davidovich-Haroche experiment is equivalent to a
non-ideal measurement of cavity $C$ photon number. Also with respect to
measurement of 
the initial state of the atom this equivalence turns out to hold. 
In this respect the third arrangement, in which the photon
number measurement is replaced by a measurement yielding also phase
information, is shown to be superior.

An interesting aspect of the generalized measurements considered here,
is that the measured probability distributions dependent on the initial
phase of the microwave fields. This makes an experimental realization
particularly challenging.


\begin{thebibliography}{10}

\bibitem{BrHa96}
{M. Brune, E. Hagley, J. Dreyer, X. Ma\^{i}tre, A. Maali, C. Wunderlich, J. M.
  Raimond and S. Haroche, {\em Phys. Rev. Let.} {\bf 77}, 4887 (1996).}

\bibitem{povm}
{E. B. Davies, {\em Quantum Theory of Open Systems}, Academic Press, London,
  1976; A. S. Holevo, {\em Probabilistic and Statistical Aspects of Quantum
  Theory}, North--Holland, Amsterdam, 1982; G. Ludwig, {\em Foundations of
  Quantum Mechanics}, Springer, Berlin, 1983, Vols. I and II; P. Busch, M.
  Grabowski and P. J. Lahti, {\em Operational quantum mechanics},
  Springer-Verlag, Berlin, Heidelberg, 1995.}

\bibitem{dMquantph99}
{W. M. de Muynck, to be published in Foundations of Physics, e-print archive
  quant-ph/9901010.}

\bibitem{StTaCoWa95}
{E.P. Storey, S.M. Tan, M.J. Collett and D.F. Walls, {\em Nature} {\bf 375},
  368 (1995).}

\bibitem{ScEnWa95}
{M.O. Scully, B.-G. Englert and H. Walther, {\em Nature} {\bf 351}, 111 (1991);
  B.-G. Englert, M.O. Scully and H. Walther, {\em Nature} {\bf 375}, 367
  (1995).}

\bibitem{DuNoRe98}
{S. Duerr, T. Nonn and G. Rempe, {\em Nature} {\bf 395}, 33 (1998).}

\bibitem{Bal70}
{L.E. Ballentine, {\em Rev. Mod. Phys.} {\bf 42}, 358 (1970).}

\bibitem{MadM90}
{H. Martens and W. de Muynck, {\em Found. of Phys.} {\bf 20}, 255, 357 (1990).}

\bibitem{WaCa84}
{N.G. Walker and J.E. Caroll, {\em Electr. Lett.} {\bf 20}, 981 (1984).}

\bibitem{YuSha80}
{H.P. Yuen, J.H. Shapiro, {\em IEEE Trans. Inform. Theory} {\bf IT--26}, 78
  (1980).}

\bibitem{SuRaTu}
{J. Summhammer, H. Rauch, and D. Tuppinger, {\em Phys. Rev. A} {\bf 36}, 4447
  (1987).}

\bibitem{MadM93}
{H. Martens and W.M. de Muynck, {\em Journ. Phys. A: Math. Gen.} {\bf 26}, 2001
  (1993).}

\bibitem{dMStMa}
{W.M. de Muynck, W.W. Stoffels, and H. Martens {\em Physica} {\bf B 175}, 127
  (1991).}

\bibitem{Ramsey}
{Norman F. Ramsey, {\em Molecular Beams}, Oxford at the Clarendon Press, First
  published 1956, Reprinted lithographically in Great Brittain from corrected
  sheets of the first edition 1963, 1969.}

\bibitem{MuMa90}
{W.M. de Muynck and H. Martens, {\em Phys. Rev. A} {\bf 42}, 5079 (1990).}

\bibitem{LePfMo98}
{D. Leibfried, T. Pfau and C.Monroe, {\em Physics Today}, april 1998, p. 22.}

\bibitem{Ba98}
{K. Banaszek, e-print archive quant-ph/9804050.}

\bibitem{HeMi96}
{C. D'Helon, G.J. Milburn, {\em Phys. Rev. A} {\bf 54}, R25 (1996).}

\bibitem{LePaAr95}
{U. Leonhardt, H. Paul and G.M. d'Ariano, {\em Phys. Rev. A} {\bf 52},4899
  (1995).}

\bibitem{VoRi89}
{K. Vogel and H. Risken, {\em Phys. Rev. A} {\bf 40}, 2847 (1989).}

\bibitem{dM98}
{W.M. de Muynck. {\em Journ. Phys. A: Math. Gen.} {\bf 31}, 431 (1998).}

\bibitem{DaHa96}
{L. Davidovich, M. Brune, J. M. Raimond and S. Haroche, {\em Phys. Rev. A} {\bf
  53}, 1295 (1996).}

\bibitem{McEl77}
{R. McEliece, {\em The theory of information and coding}, Addison--Wesley,
  London, 1977.}

\bibitem{Deutsch}
{D. Deutsch, {\em Phys. Rev. Lett.} {\bf 50}, 631 (1983).}

\bibitem{Partovi}
{M.H. Partovi, {\em Phys. Rev. Lett.} {\bf 50}, 1883 (1983).}

\bibitem{Kraus87}
{K. Kraus, {\em Phys. Rev. D} {\bf 35}, 3070 (1987).}

\bibitem{MaUf88}
{H. Maassen and J.B.M. Uffink, {\em Phys. Rev. Lett.} {\bf 60}, 1103 (1988).}

\bibitem{Paul91}
{H. Paul, {\em Quant. Opt.} {\bf 3}, 169 (1991).}

\bibitem{ViToMi98}
{D. Vitali, P. Tombesi and G.J. Milburn, {\em Phys. Rev. A} {\bf 57}, 4930
  (1998).}

\bibitem{Qfunct}
{Y. Lai and H.A. Haus, {\em Quant. Opt.} {\bf 1}, 99 (1989); M. Freyberger and
  W. Schleich, {\em Phys. Rev. A} {\bf 47}, R30 (1993); U. Leonhardt and H.
  Paul, {\em Phys. Rev. A} {\bf 47}, R2460 (1993).}

\end{thebibliography}

\section*{Acknowledgment}

The authors thank Maarten Jansen for his contribution to the 
calculations.

\newpage
\begin{center}
{\bf APPENDIX: POSITIVITY OF 
$\protect{\hat{\rho}_{\{\hat{M}_{m}\}}}$ IN THE TWO-DIMENSIONAL CASE}
\end{center}

In this appendix we prove that on a two-dimensional Hilbert space the
operator $\hat{\rho}_{\{ \hat{M}_{m}\} }$, obtained
by projecting 
density operator $\hat{\rho} $ according to (\ref{rho_projected}), is a
non-negative operator. We assume that the elements of POVM $\{
\hat{M}_{m}\} $ are linearly independent and consider the non-trivial
situation $\{ \hat{M}_{m}\} \neq \{ \hat{I}\} $
(if $\{ 
\hat{M}_{m}\} $ is uninformative, i.e. $\{
\hat{M}_{m}\} =\{ 
\hat{I}\} ,$ we get $\hat{\rho} _{\{ \hat{M}_{m}\}
}=\frac{1}{2}\hat{I}>\hat{O}$). Then 
the dimension of the subspace ${\cal H}_{\{
\hat{M}_{m}\} }$ spanned by 
the elements of $\{ \hat{M}_{m}\} $ is greater than
$1,$ and it is easy 
to prove that ${\cal H}_{\{ \hat{M}_{m}\} }$ contains
a maximal PVM $%
\{ \hat{P}_{n}\} $. Since the subspace ${\cal
H}_{\{\hat{P}_{n}\} 
} $ is a subspace of ${\cal H}_{\{\hat{M}_{m}\} }, $
the orthogonal 
projections ${\cal P}_{\{ \hat{P}_{n}\} }$ and ${\cal P}%
_{\{\hat{M}_{m}\} }$ satisfy ${\cal P}_{\{
\hat{P}_{n}\} }{\cal P}%
_{\{\hat{M}_{m}\} }={\cal P}_{\{
\hat{P}_{n}\} }$, which implies $%
Tr\hat{\rho}_{\{ \hat{M}_{m}\}
}\hat{P}_{n}=Tr\hat{\rho} \hat{P}_{n}$. So in the $\{ 
\hat{P}_{n}\} $-representation we get 
\[
\hat{\rho} =\left( 
\begin{array}{cc}
p & q \\ 
q^{\ast } & 1-p
\end{array}
\right) \text{ ; }\hat{\rho}_{\{ \hat{M}_{m}\} }=\left( 
\begin{array}{cc}
p & r \\ 
r^{\ast } & 1-p
\end{array}
\right) \text{.} 
\]

Because of the fact that ${\cal P}_{\{ \hat{M}_{m}\} }$ is an
orthogonal projection onto ${\cal H}_{\{ \hat{M}_{m}\} }$ we should
also have $Tr\hat{\rho}_{\{ \hat{M}_{m}\}
}\hat{\rho}_{\{ \hat{M}_{m}\} 
}^{\perp }=0$, with $\hat{\rho}_{\{ \hat{M}_{m}\} }^{\perp }=\left( {\cal I}-%
{\cal P}_{\{ \hat{M}_{m}\} }\right) \hat{\rho} $, implying $%
\mathop{\rm Re}%
\left( q^{\ast }r\right) -\left| r\right| ^{2}=0$, and hence 
\begin{equation}
\left| q\right| \geq \left| r\right| \text{.}  \label{q>r}
\end{equation}
Denoting the eigenvalues of $\hat{\rho}_{\{
\hat{M}_{m}\} }$ by $\lambda 
_{1} $ and $\lambda _{2}$, we find $\lambda _{1}\lambda _{2}=p\left(
1-p\right) -\left| r\right| ^{2}$. We already know that $\lambda
_{1}+\lambda _{2}=1$. So both eigenvalues are non-negative if 
\begin{equation}
p\left( 1-p\right) -\left| r\right| ^{2}\geq 0.  \label{condition_positivity}
\end{equation}
But since $\hat{\rho} >\hat{O}$ implies $p\left( 1-p\right)
-\left| q\right| ^{2}>0$ it 
directly follows from (\ref{q>r}) that condition (\ref{condition_positivity}%
) is satisfied. Hence in the two-dimensional case $\hat{\rho} _{\{
\hat{M}_{m}\} }$ is a non-negative operator.\\

\newpage
\begin{figure}[tbp]
\caption{The Haroche-Ramsey experiment.}
\label{fig1}
\end{figure}

\begin{figure}[tbp]
\caption{Comparison of non-idealities
$J^{\overline{HR}}$ and $J^{\overline{DH}}$ of the Haroche-Ramsey and
Davidovich-Haroche experiments.}
\label{fig2}
\end{figure}

\begin{figure}[tbp]
\caption{Plots of a) $J_{(\protect\lambda )}$ and b) $J_{(\protect\mu
)}$ as functions of $\protect\gamma $ and $\protect\Phi$.}
\label{fig3}
\end{figure}

\begin{figure}[tbp]
\caption{Plot of $J_{(\protect\lambda) }$ versus $J_{(\protect\mu) }$
for $\protect\delta =0, \protect\Phi =\frac{\protect\pi }{2}$, demonstrating
complementarity of interference and path observables. The straight
line represents the lower bound of $J_{(\protect\lambda)}
+J_{(\protect\mu) }$, given by inequality (\protect\ref{3}).}
\label{fig4}
\end{figure}

\begin{figure}[tbp]
\caption{Plots of a) $J_{(\protect\lambda')}$ and b) $J_{(\protect\mu'
)}$ as functions of $\protect\gamma sin\Phi$ and $\protect\delta$.}
\label{fig5}
\end{figure}

\newpage
\begin{figure}[t]
\leavevmode
\centerline{
\begin{picture}(300,825)(0,0)
 \epsfysize=1.5in
 \put (20,675){\epsfbox{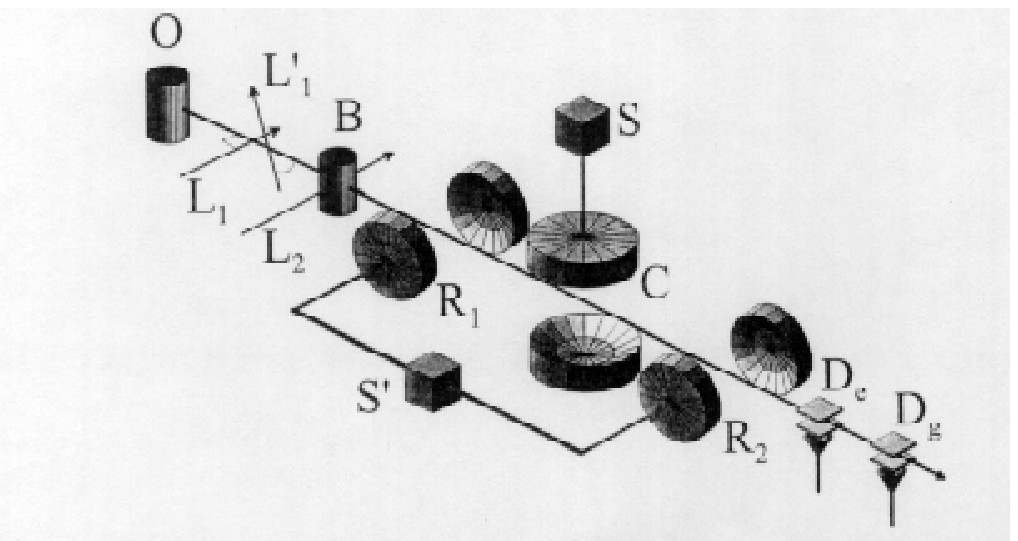}} 
\put (100,650){Figure 1}
 \epsfysize=1.2in
 \put (0,195){\epsfbox{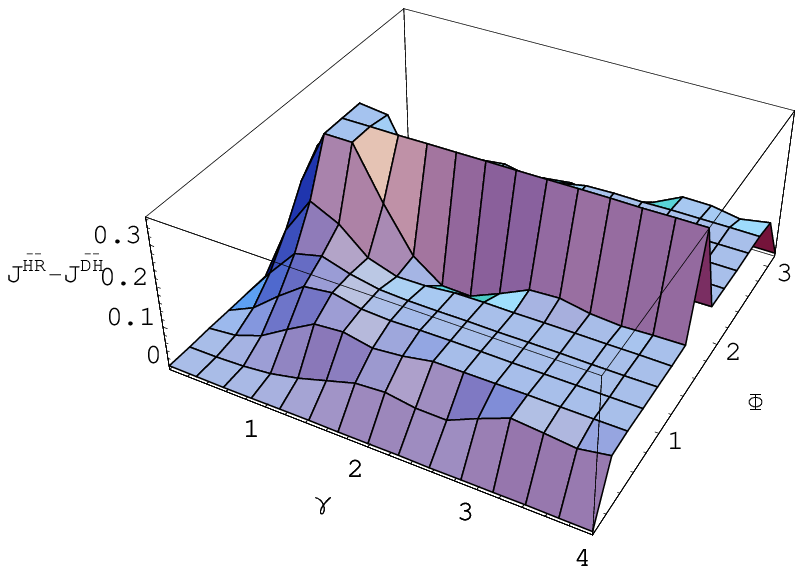}} 
\put (100,150){Figure 2}
\end{picture}
}
\end{figure}

\newpage
\begin{figure}[t]
\leavevmode
\centerline{
\begin{picture}(300,800)(0,0)
 \epsfysize=1.5in
 \put (50,550){\epsfbox{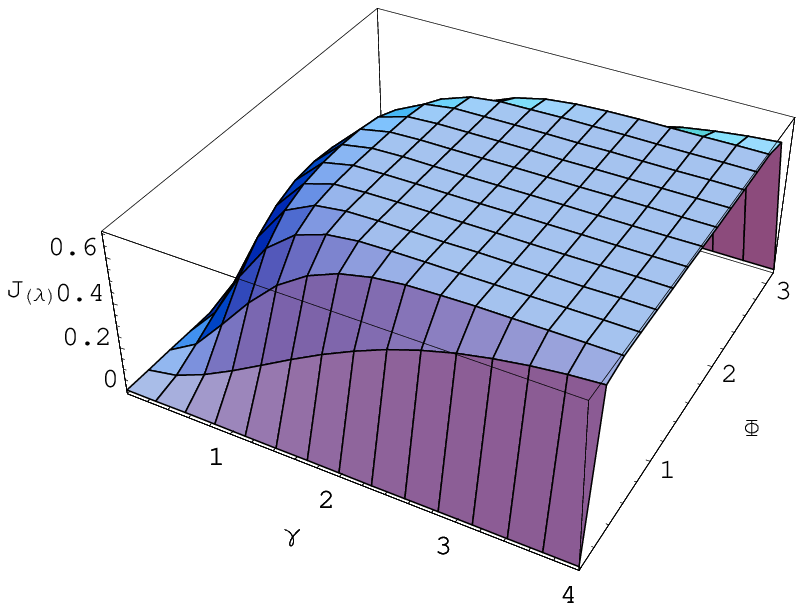}} 
\put (-100,750){a)}
 \epsfysize=1.5in
 \put (50,230){\epsfbox{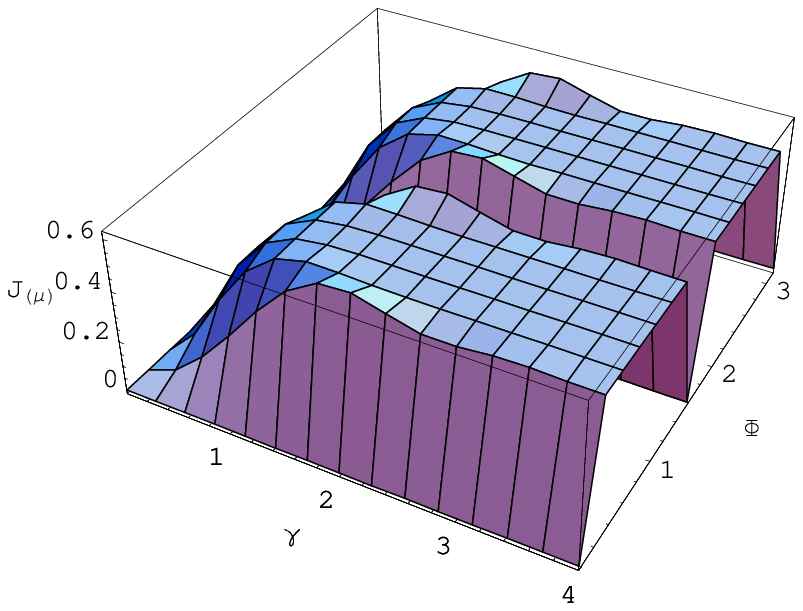}} 
\put (-100,450){b)}
\put (100,150){Figure 3}
\end{picture}
}
\end{figure}

\newpage
\begin{figure}[t]
\leavevmode
\centerline{
\begin{picture}(300,825)(0,0)
 \epsfysize=2.7in
 \put (80,525){\epsfbox{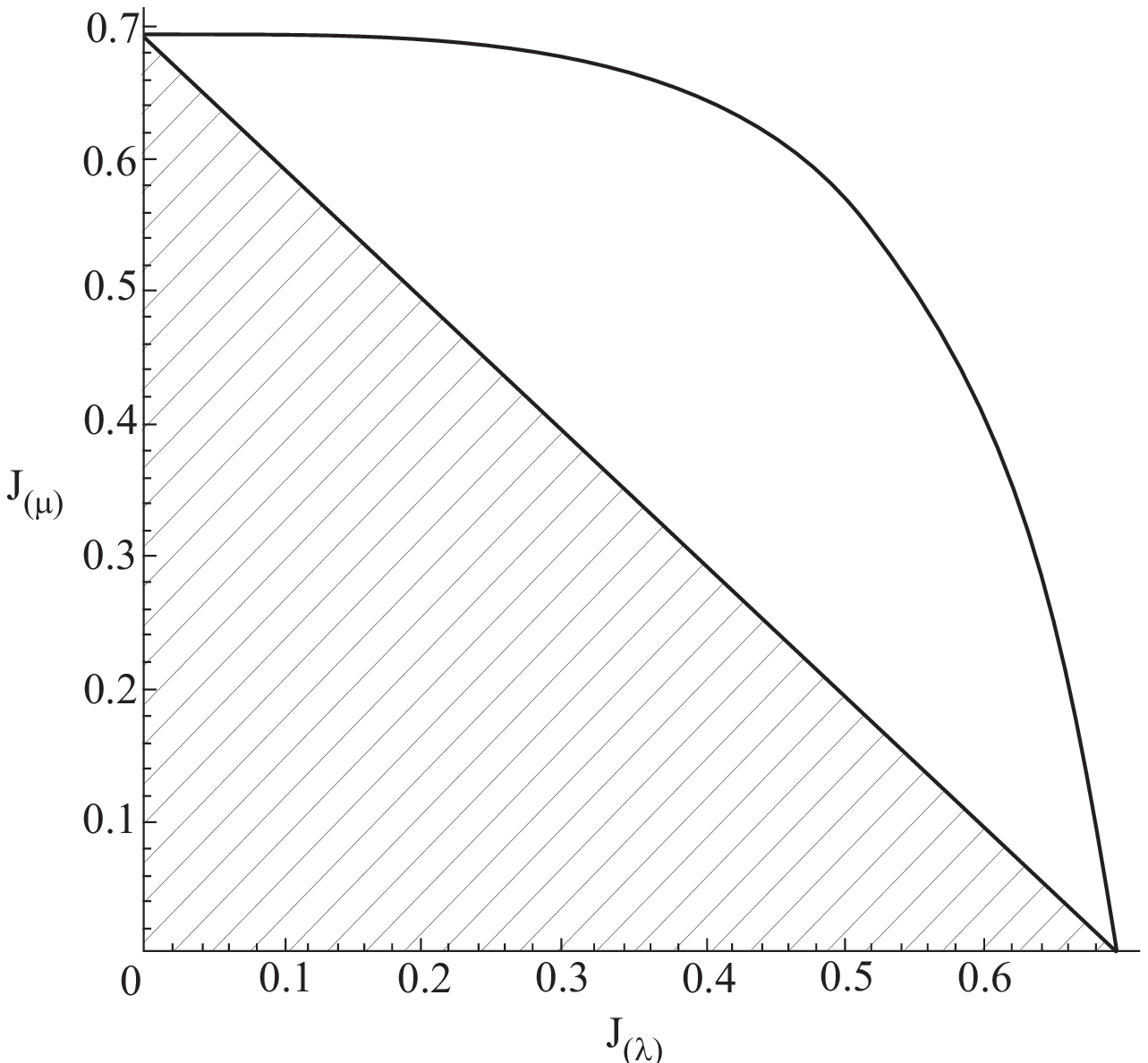}} 
\put (100,450){Figure 4.}
\end{picture}
}
\end{figure}

\newpage
\begin{figure}[t]
\leavevmode
\centerline{
\begin{picture}(300,800)(0,0)
 \epsfysize=1.5in
 \put (100,580){\epsfbox{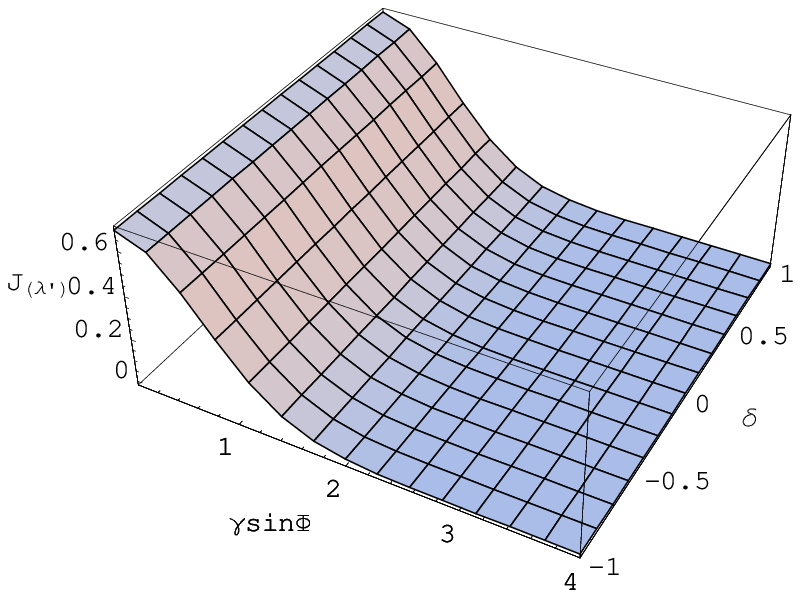}} 
\put (-75,820){a)}
 \epsfysize=1.5in
 \put (100,250){\epsfbox{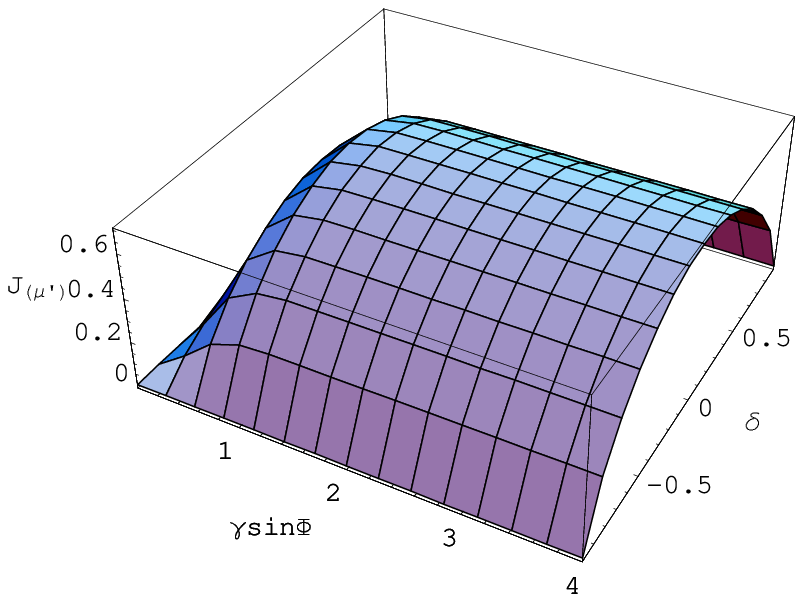}} 
\put (-75,520){b)}
\put (100,150){Figure 5.}
\end{picture}
}
\end{figure}

\end{document}